\documentclass[a4paper,11pt]{article}
\pdfoutput=1 
\usepackage{cmbright}
\usepackage{color, colortbl}
\usepackage[utf8]{inputenc} 
\usepackage[T1]{fontenc}    
\usepackage{url}            
\usepackage{booktabs}       
\usepackage{amsfonts}       
\usepackage[centertags]{amsmath}
\usepackage{amssymb,amscd,amsbsy}
\usepackage{nicefrac}       
\usepackage{microtype}      
\usepackage{xcolor}
\usepackage[alsoload=hep, separate-uncertainty = true]{siunitx}
\usepackage{lmodern}
\usepackage{graphicx}
\usepackage[babel]{csquotes}\label{key}
\usepackage{tabularx}
\usepackage{appendix}

\usepackage{arydshln}
\usepackage{multirow}
\usepackage{hhline}
\usepackage{fancyhdr}
\usepackage{etoolbox}
\usepackage[percent]{overpic}

\graphicspath{ {img/} }

\definecolor{Gray}{gray}{0.9}
\definecolor{LightRed}{rgb}{1, 0.92, 0.92}

\DeclareSIUnit\year{y}
\newunit{\evts}{evt}

\newcommand\Tstrut{\rule{0pt}{2.6ex}}         
\newcommand\Bstrut{\rule[-0.9ex]{0pt}{0pt}}   

\usepackage{jinstpub} 
\providecommand{\autocite}{\cite}

\title{Event reconstruction for KM3NeT/ORCA using convolutional neural networks} 

\author[a]{S.~Aiello,}
\author[bb,b]{A.~Albert,}
\author[c]{S. Alves Garre,}
\author[d]{Z.~Aly,}
\author[e]{F.~Ameli,}
\author[f]{M.~Andre,}
\author[g]{G.~Androulakis,}
\author[h]{M.~Anghinolfi,}
\author[i]{M.~Anguita,}
\author[j]{G.~Anton,}
% ...
\affiliation[a]{INFN, Sezione di Catania, Via Santa Sofia 64, Catania, 95123 Italy}
\affiliation[b]{IN2P3, IPHC, 23 rue du Loess, Strasbourg, 67037 France}
\affiliation[c]{IFIC - Instituto de F{\'\i}sica Corpuscular (CSIC - Universitat de Val{\`e}ncia), c/Catedr{\'a}tico Jos{\'e} Beltr{\'a}n, 2, 46980 Paterna, Valencia, Spain}
\affiliation[d]{Aix~Marseille~Univ,~CNRS/IN2P3,~CPPM,~Marseille,~France}
\affiliation[e]{INFN, Sezione di Roma, Piazzale Aldo Moro 2, Roma, 00185 Italy}
\affiliation[f]{Universitat Polit{\`e}cnica de Catalunya, Laboratori d'Aplicacions Bioac{\'u}stiques, Centre Tecnol{\`o}gic de Vilanova i la Geltr{\'u}, Avda. Rambla Exposici{\'o}, s/n, Vilanova i la Geltr{\'u}, 08800 Spain}
\affiliation[g]{NCSR Demokritos, Institute of Nuclear and Particle Physics, Ag. Paraskevi Attikis, Athens, 15310 Greece}
\affiliation[h]{INFN, Sezione di Genova, Via Dodecaneso 33, Genova, 16146 Italy}
\affiliation[i]{University of Granada, Dept.~of Computer Architecture and Technology/CITIC, 18071 Granada, Spain}
\affiliation[j]{Friedrich-Alexander-Universit{\"a}t Erlangen-N{\"u}rnberg, Erlangen Centre for Astroparticle Physics, Erwin-Rommel-Stra{\ss}e 1, 91058 Erlangen, Germany}
% ...
\affiliation[bb]{Universit{\'e} de Strasbourg, Universit{\'e} de Haute Alsace, GRPHE, 34, Rue du Grillenbreit, Colmar, 68008 France}
% ...

\emailAdd{michael.m.moser@fau.de}
\emailAdd{thomas.eberl@fau.de}


\abstract{

The KM3NeT research infrastructure is currently under construction at
two locations in the Mediterranean Sea. The KM3NeT/ORCA
water-Cherenkov neutrino detector off the French coast will instrument
several megatons of seawater with photosensors. Its main objective is
the determination of the neutrino mass ordering.
This work aims at demonstrating the general applicability of
deep convolutional neural networks to neutrino telescopes, using
simulated datasets for the KM3NeT/ORCA detector as an example.
To this end, the networks are employed to achieve reconstruction and
classification tasks that constitute an alternative to the analysis
pipeline presented for KM3NeT/ORCA in the KM3NeT Letter of Intent.
They are used to infer
event reconstruction estimates for the energy, the direction, and the
interaction point of incident neutrinos.
The spatial distribution of Cherenkov light generated by charged
particles induced in neutrino interactions is classified as shower- or
track-like, and the main background processes associated with the
detection of atmospheric neutrinos are recognized.
Performance comparisons to machine-learning classification and
maximum-likelihood reconstruction algorithms previously developed for
KM3NeT/ORCA are provided.
It is shown that this application of
deep convolutional neural networks to simulated datasets for a
large-volume neutrino telescope yields competitive reconstruction
results and performance improvements with respect to classical
approaches.

}

\DeclareSymbolFont{EUr}{U}{eur}{m}{n}
\DeclareSymbolFont{EUb}{U}{eur}{b}{n}
\DeclareMathSymbol{\upGamma}{\mathord}{EUr}{"00}
\DeclareMathSymbol{\upDelta}{\mathord}{EUr}{"01}
\DeclareMathSymbol{\upTheta}{\mathord}{EUr}{"02}
\DeclareMathSymbol{\upLambda}{\mathord}{EUr}{"03}
\DeclareMathSymbol{\upXi}{\mathord}{EUr}{"04}
\DeclareMathSymbol{\upPi}{\mathord}{EUr}{"05}
\DeclareMathSymbol{\upSigma}{\mathord}{EUr}{"06}
\DeclareMathSymbol{\upUpsilon}{\mathord}{EUr}{"07}
\DeclareMathSymbol{\upPhi}{\mathord}{EUr}{"08}
\DeclareMathSymbol{\upPsi}{\mathord}{EUr}{"09}
\DeclareMathSymbol{\upOmega}{\mathord}{EUr}{"0A}
\DeclareMathSymbol{\upalpha}{\mathord}{EUr}{"0B}
\DeclareMathSymbol{\upbeta}{\mathord}{EUr}{"0C}
\DeclareMathSymbol{\upgamma}{\mathord}{EUr}{"0D}
\DeclareMathSymbol{\updelta}{\mathord}{EUr}{"0E}
\DeclareMathSymbol{\upepsilon}{\mathord}{EUr}{"0F}
\DeclareMathSymbol{\upzeta}{\mathord}{EUr}{"10}
\DeclareMathSymbol{\upeta}{\mathord}{EUr}{"11}
\DeclareMathSymbol{\uptheta}{\mathord}{EUr}{"12}
\DeclareMathSymbol{\upiota}{\mathord}{EUr}{"13}
\DeclareMathSymbol{\upkappa}{\mathord}{EUr}{"14}
\DeclareMathSymbol{\uplambda}{\mathord}{EUr}{"15}
\DeclareMathSymbol{\upmu}{\mathord}{EUr}{"16}
\DeclareMathSymbol{\upnu}{\mathord}{EUr}{"17}
\DeclareMathSymbol{\upxi}{\mathord}{EUr}{"18}
\DeclareMathSymbol{\uppi}{\mathord}{EUr}{"19}
\DeclareMathSymbol{\uprho}{\mathord}{EUr}{"1A}
\DeclareMathSymbol{\upsigma}{\mathord}{EUr}{"1B}
\DeclareMathSymbol{\uptau}{\mathord}{EUr}{"1C}
\DeclareMathSymbol{\upupsilon}{\mathord}{EUr}{"1D}
\DeclareMathSymbol{\upphi}{\mathord}{EUr}{"1E}
\DeclareMathSymbol{\upchi}{\mathord}{EUr}{"1F}
\DeclareMathSymbol{\uppsi}{\mathord}{EUr}{"20}
\DeclareMathSymbol{\upomega}{\mathord}{EUr}{"21}
\DeclareMathSymbol{\upvarepsilon}{\mathord}{EUr}{"22}
\DeclareMathSymbol{\upvartheta}{\mathord}{EUr}{"23}
\DeclareMathSymbol{\upvaromega}{\mathord}{EUr}{"24}
\DeclareMathSymbol{\upvarphi}{\mathord}{EUr}{"27}
\DeclareMathSymbol{\UpGamma}{\mathord}{EUb}{"00}
\DeclareMathSymbol{\UpDelta}{\mathord}{EUb}{"01}
\DeclareMathSymbol{\UpTheta}{\mathord}{EUb}{"02}
\DeclareMathSymbol{\UpLambda}{\mathord}{EUb}{"03}
\DeclareMathSymbol{\UpXi}{\mathord}{EUb}{"04}
\DeclareMathSymbol{\UpPi}{\mathord}{EUb}{"05}
\DeclareMathSymbol{\UpSigma}{\mathord}{EUb}{"06}
\DeclareMathSymbol{\UpUpsilon}{\mathord}{EUb}{"07}
\DeclareMathSymbol{\UpPhi}{\mathord}{EUb}{"08}
\DeclareMathSymbol{\UpPsi}{\mathord}{EUb}{"09}
\DeclareMathSymbol{\UpOmega}{\mathord}{EUb}{"0A}
\DeclareMathSymbol{\Upalpha}{\mathord}{EUb}{"0B}
\DeclareMathSymbol{\Upbeta}{\mathord}{EUb}{"0C}
\DeclareMathSymbol{\Upgamma}{\mathord}{EUb}{"0D}
\DeclareMathSymbol{\Updelta}{\mathord}{EUb}{"0E}
\DeclareMathSymbol{\Upepsilon}{\mathord}{EUb}{"0F}
\DeclareMathSymbol{\Upzeta}{\mathord}{EUb}{"10}
\DeclareMathSymbol{\Upeta}{\mathord}{EUb}{"11}
\DeclareMathSymbol{\Uptheta}{\mathord}{EUb}{"12}
\DeclareMathSymbol{\Upiota}{\mathord}{EUb}{"13}
\DeclareMathSymbol{\Upkappa}{\mathord}{EUb}{"14}
\DeclareMathSymbol{\Uplambda}{\mathord}{EUb}{"15}
\DeclareMathSymbol{\Upmu}{\mathord}{EUb}{"16}
\DeclareMathSymbol{\Upnu}{\mathord}{EUb}{"17}
\DeclareMathSymbol{\Upxi}{\mathord}{EUb}{"18}
\DeclareMathSymbol{\Uppi}{\mathord}{EUb}{"19}
\DeclareMathSymbol{\Uprho}{\mathord}{EUb}{"1A}
\DeclareMathSymbol{\Upsigma}{\mathord}{EUb}{"1B}
\DeclareMathSymbol{\Uptau}{\mathord}{EUb}{"1C}
\DeclareMathSymbol{\Upupsilon}{\mathord}{EUb}{"1D}
\DeclareMathSymbol{\Upphi}{\mathord}{EUb}{"1E}
\DeclareMathSymbol{\Upchi}{\mathord}{EUb}{"1F}
\DeclareMathSymbol{\Uppsi}{\mathord}{EUb}{"20}
\DeclareMathSymbol{\Upomega}{\mathord}{EUb}{"21}
\DeclareMathSymbol{\Upvarepsilon}{\mathord}{EUb}{"22}
\DeclareMathSymbol{\Upvartheta}{\mathord}{EUb}{"23}
\DeclareMathSymbol{\Upvaromega}{\mathord}{EUb}{"24}
\DeclareMathSymbol{\Upvarphi}{\mathord}{EUb}{"27}
\catcode`\@=11 
\def\parenbar{\mathpalette\p@renb@r}
\def\p@renb@r#1#2{\vbox{%
  \ifx#1\scriptscriptstyle \dimen@.7em\dimen@ii.2em\else
  \ifx#1\scriptstyle \dimen@.8em\dimen@ii.25em\else
  \dimen@1em\dimen@ii.4em\fi\fi \offinterlineskip
  \ialign{\hfill##\hfill\cr
    \vbox{\hrule width\dimen@ii}\cr
    \noalign{\vskip-.3ex}%
    \hbox to\dimen@{$\mathchar300\hfil\mathchar301$}\cr
    \noalign{\vskip-.3ex}%
    $#1#2$\cr}}}
\catcode`\@=12 
\def\nuan{\parenbar{\upnu}\kern-0.4ex}
\def\ean{\parenbar{\text{e}}\kern-0.4ex}
\def\muan{\parenbar{\upmu}\kern-0.4ex}
\def\tauan{\parenbar{\uptau}\kern-0.4ex}
\newcommand{\overbar}[1]{\mkern 1.5mu\overline{\mkern-1.5mu#1\mkern-1.5mu}\mkern 1.5mu}
\newcommand{\anu}{\overbar{\upnu}}
\newcommand{\nue}{\upnu_{\text{e}}}
\newcommand{\numu}{\upnu_{\upmu}}
\newcommand{\nutau}{\upnu_{\uptau}}
\newcommand{\nueCC}{\upnu_{\text{e}}^{\text{CC}}}
\newcommand{\numuCC}{\upnu_{\upmu}^{\text{CC}}}
\newcommand{\nutauCC}{\upnu_{\uptau}^{\text{CC}}}
\newcommand{\nuCC}{\upnu^{\text{CC}}}

\newcommand{\nueNC}{\upnu_{\text{e}}^{\text{NC}}}
\newcommand{\nueCCNC}{\upnu_{\text{e}}^{\text{CC/NC}}}
\newcommand{\anueCC}{\overbar{\upnu}_{\text{e}}^{\text{CC}}}
\newcommand{\anumuCC}{\overbar{\upnu}_{\upmu}^{\text{CC}}}
\newcommand{\anuCC}{\overbar{\upnu}^{\text{CC}}}

\newcommand{\nuanmuCC}{\nuan_{\upmu}^{\;{}\text{CC}}}

\newcommand\mysref[1]{Sec.\;\ref{#1}}
\newcommand\myfref[1]{Fig.\;\ref{#1}}
\newcommand\myeref[1]{Eq.\;\ref{#1}}
\newcommand\mytref[1]{Tab.\;\ref{#1}}
\newcommand{\yrecovec}{\vec{\upxi}_{\text{reco}}}
\newcommand{\yreco}{\upxi_{\text{reco}}}
\newcommand{\ytrue}{\upxi_{\text{true}}}
\newcommand{\kforty}{{^{\text{40}}}\text{K}}

\begin{document}
\maketitle
\flushbottom

\newpage
\clearpage

\section{Introduction}

Precision measurements of the fundamental properties of neutrinos are
one of the opportunities that might allow us to discover and
understand the physics that exists beyond the established Standard
Model of particle physics.

The detection of neutrinos, both for fundamental particle physics and
high-energy astrophysics, can be achieved with the deep-sea and
photon-detection technology that has been developed by the ANTARES
\autocite{Collaboration:2011nsa} and KM3NeT
\autocite{Adrian-Martinez2016} Collaborations for very-large-volume
water-Cherenkov detectors.

KM3NeT/ORCA, the low-energy detector of KM3NeT, addresses the
determination of a still unknown, but fundamental parameter of
neutrino physics: the neutrino mass ordering. The experiment focuses
on the measurement of the energy- and zenith-angle-dependent
oscillation patterns of cosmic-ray-induced neutrinos with a few-GeV
energy that originate in the atmosphere and traverse the Earth
\autocite{2013JHEP...02..082A}.

The power to distinguish between the two different mass orderings is
linked to the detection of an excess or deficit of neutrino events in
different regions of these oscillation patterns.
This sensitivity
increases with better energy and zenith-angle resolution and flavour
identification for the interacting neutrinos, and finer control of
systematic effects that influence the measurement. Therefore, one of
the most important goals in the analysis of KM3NeT/ORCA data is the
development and characterisation of neutrino event reconstruction and
classification algorithms that improve these resolutions.

The neutrino detection principle of water- or ice-based large-scale
Cherenkov detectors relies on the detection of Cherenkov photons
induced by charged secondary particles created in a neutrino
interaction with the target material.
All neutrino flavours can interact through the weak neutral current
(NC) mediated by the exchange of a $\text{Z}^0$ boson. This
interaction results in a particle shower composed mainly of hadrons,
generically referred to as a \textit{hadronic system}, while the scattered
neutrino escapes undetected.
An interaction via the weak charged current (CC), with the exchange of
a $\text{W}^+$ or $\text{W}^-$ boson, also often results in a hadronic
shower at the interaction vertex.  Additionally, a lepton of the same
flavour as the interacting neutrino is created, which carries a fraction
of the incoming neutrino energy.


A muon neutrino or muon antineutrino CC interaction,
$\nuanmuCC$\footnote{The symbol $\nuan$ denotes both, neutrino and
  antineutrino, at the same time.}, results in an outgoing muon in the
final state. From now on the term `neutrino' refers always to both
neutrinos and antineutrinos, if not stated otherwise.
The muon appears as a track-like light source in the detector, and can
therefore be identified with good confidence, depending on its track
length.
The visible trajectory of the muon is determined by its energy loss,
and in water it amounts to about roughly 4\,m per GeV of muon energy
for the relevant energy regime of a few GeV.

At the energy ranges considered in KM3NeT/ORCA, all neutrino-nucleon
NC, $\nueCC$, and $\nutauCC$ interactions, with the exception of
roughly 18\,\% of tau leptons decaying into muons, create a particle
shower of a few meters length, that appears as an elongated, but
localised, light source compared to the typical distance scales
between the detector elements (\SIrange[range-units=single,
  range-phrase=--]{9}{20}{\metre},
cf.~\mysref{subsect:detector_layout}).  This event type is referred to
as shower-like.  The outgoing electron from a $\nueCC$ event initiates
an electromagnetic shower, a cascade of $\text{e}^\pm$-pairs, while
the hadronic system, typically at the neutrino interaction vertex,
develops into a hadronic shower with large event-to-event fluctuations
and a possibly complex structure of hadronic or electromagnetic
sub-showers, depending on the decay modes of individual particles in
the shower.

Although an electromagnetic shower consists of many
$\text{e}^\pm$-pairs with rather short path lengths (about 36\,cm radiation
length in water, see Sec.~33.4.2. in Ref.~\autocite{Tanabashi:2018oca}) and
overlapping Cherenkov cones, the small pair opening angle preserves
the Cherenkov angle peak of the total angular light distribution. This
results in a single Cherenkov ring projected onto the plane
perpendicular to the shower axis.  Similarly, each hadronic shower
particle with energy above the Cherenkov threshold will produce a
Cherenkov ring. Therefore, hadronic showers show a variety of
different signatures due to the various possible combinations of
initial hadron types, their momenta and the diversity of their
hadronic interactions in the shower evolution.

While electromagnetic cascades show only negligible fluctuations in
the number of emitted Cherenkov photons and in the angular light
distribution, hadronic cascades show significant intrinsic
fluctuations in the relevant few-GeV energy range.  These intrinsic
fluctuations of hadronic cascades and the resulting limitations for
the energy and angular resolutions have been studied in detail in
Ref.~\autocite{Adrian-Martinez:2016zzs}.

Dedicated reconstruction algorithms for track-like and shower-like
events have been developed for KM3NeT/ORCA based on maximum-likelihood
methods. Additionally, a machine-learning algorithm, based on Random
Forests \autocite{Breiman2001}, has been employed successfully to
classify track-like, shower-like, and background events. These
algorithms, their implementation and performance are described in the
KM3NeT Letter of Intent \autocite{Adrian-Martinez2016}.

In the last few years, significant progress has been made in the
machine-learning community due to the advent of deep-learning
techniques.  A particularly successful deep-learning concept is that
of a deep neural network. Specialised neural network model
architectures have been designed for individual use cases.  In the
field of computer vision, Convolutional Neural Networks (CNNs) have
led to a strong increase in image recognition performance. From 2010
to 2016, the error rates in e.g.\ the popular ImageNet image
classification challenge improved by a factor of 10
\autocite{Russakovsky2014}\autocite{ilsvrc2017_talk}.

Since the data of many high-energy physics experiments can be
interpreted in a way similar to typical images in the computer vision
domain, these techniques have already been exploited by several
experiments
\autocite{Huennefeld:2017pdh,Erdmann:2017str,Shilon:2018xlp,Guest:2018yhq}.
As an example, the classification performance of neutrino interactions
in the NOvA experiment has been significantly improved by employing
CNNs compared to classical reconstruction tools
\autocite{Aurisano2016}.

In this paper, we present for the first time the application of CNNs
to detailed Monte Carlo simulations of a large water-Cherenkov
neutrino detector with the goal to provide a comprehensive
reconstruction pipeline for KM3NeT/ORCA, starting from data at the
level of the data-acquisition system.
For this purpose,
a Keras-based \autocite{chollet2015keras} software framework, called
\textit{OrcaNet} \autocite{orcanet}, has been developed that
simplifies the usage of neural networks for neutrino telescopes.
Here, we apply this framework to neutrino event reconstruction and
classification in KM3NeT/ORCA, and compare the results achieved with
the algorithms described in Ref.~\autocite{Adrian-Martinez2016}.
We note that also these algorithms
continue to be developed and improved in KM3NeT.

The paper is organised as follows.  Section 2 introduces the
KM3NeT/ORCA detector, the Monte Carlo simulation chain used to
generate training and validation data for the CNNs and introduces
relevant aspects of the trigger algorithms. Section 3 gives a brief
introduction to the main functional features of the employed CNNs,
while Section 4 details the developed pre-processing chain that creates
suitable input images from the Monte Carlo simulation data. Section 5
provides an overview of the general network architecture that is
shared by all CNNs that have been designed for the reconstruction and
classification tasks, which together define the analysis pipeline for
KM3NeT/ORCA. The concepts and performance of these specific CNNs, as
well as exemplary comparisons to their counterpart algorithms, are
explained in the next sections. Section 6 explains the background
classifier,
Section 7 the event topology classifier used to
distinguish track-like and shower-like events, while Section 8
introduces event regression and its respective uncertainties, i.e. the
reconstruction of the direction, energy, and vertex of the incident
neutrinos. Section 9 summarises and concludes the paper.

\section{The KM3NeT/ORCA experiment}

The KM3NeT research infrastructure is under construction at two sites
in the Mediterranean Sea. The KM3NeT/ORCA detector is located about
\SI{40}{\kilo\metre} off-shore of Toulon in the south of France. Its
main goal is to detect atmospheric neutrinos with GeV energies
(\SIrange[range-units=single,range-phrase=--]{3}{40}{\giga\electronvolt}),
while KM3NeT/ARCA, located south-east of Sicily, aims to investigate
astrophysical neutrinos. The main design principles and scientific
goals of the experiment can be found in the KM3NeT Letter of Intent
\autocite{Adrian-Martinez2016}.

\subsection{Layout of the detector}
\label{subsect:detector_layout}
The detector volume of KM3NeT/ORCA will be instrumented with 115
Detection Units (DUs), which are vertical, string-like structures
anchored to the seabed and held upright by a buoy at the top of the
DU. Currently, the first six DUs have been installed and are
operational.  Each DU holds 18 Digital Optical Modules (DOMs). The
DOMs contain 31 photomultiplier tubes (PMTs) with a diameter of 3"
each.  The PMTs are used to measure two quantities, the arrival time
and the time range that the anode output signal remains above a
tunable threshold (time-over-threshold, ToT) with a time resolution on
the nanosecond scale.  The ToT can be used as a proxy for the amount
of light registered by the PMT.  The vertical spacing between the DOMs
on a single DU is on average \SI{9}{\metre}, while the average
horizontal distance between the DUs is about \SI{20}{\metre}. This
results in a total instrumented volume of about six megatons of
seawater.

There are two main sources of background in KM3NeT/ORCA, namely
atmospheric muons reaching the detector from above and random optical
background due to beta decays of $\kforty$ in seawater and
bioluminescence.  Optical background, which is dominated by decays of
$\kforty$, accounts for about \SI{7}{\kilo\hertz} of uncorrelated
single photon noise per PMT with a rate of two-fold coincidences of
about \SI{500}{\hertz} per DOM \autocite{Ageron2020}.

The atmospheric muon background can be reduced significantly by
requiring the reconstructed vertex position to be inside or close to
the instrumented detector volume. In addition, predominantly
atmospheric muons enter the detector from above and therefore the
atmospheric muon background can be further reduced by discarding
events for which the direction of the emerging particle trajectory is
reconstructed as downwards.


\subsection{Monte Carlo simulations and trigger algorithms}
\label{sect:simulations_and_trigger}

Detailed Monte Carlo (MC) simulations of the detector response have
been produced for three distinct types of triggered data, namely
atmospheric muons, random noise and neutrinos. A detailed introduction
to the KM3NeT simulation package and the trigger algorithms can be
found in Ref.~\autocite{Adrian-Martinez2016}.

For neutrinos, $\nueCC$ and $\numuCC$ interactions on nucleons and
nuclei in seawater have been simulated, while all NC interactions are
represented by $\nueNC$, since the detector signature is identical for
all flavours.  Charged-current interactions of $\nutau$ are neglected
for simplicity, as the resulting detector signatures for the different
decay modes of the tau lepton are very similar to either $\nueCC$ or
$\numuCC$.

The distance between DUs for the simulations employed in this work is
on average \SI{23}{\metre} and hence \SI{3}{\metre} larger than for
the simulations used in Ref.~\autocite{Adrian-Martinez2016}.
The vertical inter-DOM spacing is set to \SI{9}{\metre} on average,
which was identified as optimal for the determination of the neutrino
mass ordering in Ref.~\autocite{Adrian-Martinez2016}. The highest level
of simulated data consists of a list of \textit{hits}, i.e.\  time
stamp, ToT, and identifier, for all PMTs in the detector.
In addition to the \textit{signal}
hits induced by the interactions of neutrinos and by atmospheric muons
in the sensitive detector volume, simulated \textit{background} hits
due to random noise are added such that the simulated triggered data
matches the real conditions as closely as possible.

After hits have been simulated, several trigger algorithms that rely
on causality conditions are applied.
Once a 
trigger has fired to define an \textit{event}, all hits that have
fired the trigger, including signal and background hits,
are labelled as \textit{triggered} hits. Since the trigger algorithm
is not fully efficient in identifying all signal hits, a larger time
window than the one defined for the triggered hits is saved for
further analysis.  Assuming a triggered event with $t_{\text{first}}$
as the time of the first triggered hit and $t_{\text{last}}$ as the
time of the last triggered hit, all photon hits in each PMT in a time window
[$t_{\text{first}} - t_{\text{marg}}$, $t_{\text{last}} +
  t_{\text{marg}}$] are recorded, where $t_{\text{marg}}$ is defined
by the maximum amount of time that a photon propagating in water
would need to traverse the whole detector. As a result, the total time
window of triggered neutrino events in KM3NeT/ORCA is about
\SI{3}{\micro\second}.

A summary with detailed information about the simulated data for
KM3NeT/ORCA is shown in \mytref{tab:TabSim1}.
\setlength{\dashlinedash}{0.2pt}
\setlength{\dashlinegap}{4.5pt}
\setlength{\arrayrulewidth}{0.2pt}
\renewcommand{\arraystretch}{1.1}
\begin{table}[h!]
	\centering
	\setlength{\extrarowheight}{0.2cm}
	\sisetup{range-phrase=-}
	\sisetup{range-units=single}
	\begin{tabular}{|c|c|c|c|}
          \hline
          \rowcolor{Gray} Event type & $\text{E}_\text{gen}$ spectrum & $\text{N}_\text{trig}$ [10$^{\text{6}}$]& Energy range\Bstrut \\
          \hline
		Atmospheric muon & - & 65.2 & - \Tstrut\\ \hdashline 
		Random noise & - & 23.3 & - \Tstrut\\ \hdashline 
		$\nueNC$ & $\text{E}^{\text{-1}}$ & 1.1 & \SIrange{1}{5}{\giga\electronvolt} \Tstrut\\ \hdashline 
		$\nueNC$ & $\text{E}^{\text{-3}}$ & 3.7 & \SIrange{3}{100}{\giga\electronvolt} \Tstrut\\ \hdashline 
		$\nueCC$ & $\text{E}^{\text{-1}}$ & 1.5 & \SIrange{1}{5}{\giga\electronvolt} \Tstrut\\ \hdashline 
		$\nueCC$ & $\text{E}^{\text{-3}}$ & 4.4 & \SIrange{3}{100}{\giga\electronvolt} \Tstrut\\ \hdashline 
		$\numuCC$ & $\text{E}^{\text{-1}}$ & 1.7 & \SIrange{1}{5}{\giga\electronvolt} \Tstrut\\ \hdashline 
		$\numuCC$ & $\text{E}^{\text{-3}}$ & 8.3 & \SIrange{3}{100}{\giga\electronvolt} \Tstrut \\ \hline 
	\end{tabular}
	\vspace{5mm}
	\caption[List of available ORCA 115 line simulations.]{List of
          Monte Carlo simulations for a KM3NeT/ORCA detector composed
          of 115 DUs. The first column reports the simulated event
          type. The neutrino simulations comprise neutrino and
          antineutrino interactions of the indicated type. The second
          column specifies the power law used to simulate the energy
          spectrum of the interacting neutrinos. A reweighting scheme
          is used in this work, where appropriate, to simulate an
          atmospheric neutrino flux model
          \autocite{HKKM}. $\text{N}_\text{trig}$ is the number of
          events that remain after triggering.  Atmospheric muons have
          been simulated with the MUPAGE package
          \autocite{Carminati2009}. Random noise events have been
          simulated conservatively with a \SI{10}{\kilo\hertz} single
          rate per PMT with additional n-fold coincidences
          (\SI{600}{\hertz} two-fold, \SI{60}{\hertz} three-fold,
          \SI{7}{\hertz} four-fold, \SI{0.8}{\hertz} five-fold and
          \SI{0.08}{\hertz} six-fold
          \autocite{Ageron2020}). Time-varying increases of the hit
          rate due to bioluminescence in seawater have not been
          simulated.}
	\label{tab:TabSim1}
\end{table}
\renewcommand{\arraystretch}{1.0}

\section{Convolutional neural networks}
\label{sec:cnns}
This section introduces the concepts and nomenclature used in the
description of the networks that have been developed and used in this
work.

Convolutional neural networks \autocite{Goodfellow2016,cs231n} form a
specialised class of deep neural networks.  Generally, neural networks
are used in order to approximate a function $f(x)$, which maps a
certain number of inputs $x_i \in X$ to some outputs $y_i \in Y$.  The
goal is then to find an approximation $\hat{f}(x)$ to the function
$f(x)$ that describes the relationship between the inputs $x_i$ and
the outputs $y_i$.  Neural networks are based on the concept of
artificial neurons that are arranged in layers. For a fully
connected neural network, each neuron in a layer is connected to all
neurons in the previous layer.

Stacking multiple layers of neurons can be interpreted as multiple
functions that are acting on the input $X$ in a chain. For a two-layer
network and thus two functions $f^{(1)}$ and $f^{(2)}$ (the
~$\hat{}$~ symbol of $\hat{f}$ is neglected from this point on) this is:
\begin{equation}
f(x) = f^{(2)}(f^{(1)}(x)).
\end{equation}
Here, $f^{(1)}$ refers to the first layer in the network and $f^{(2)}$
to the second. The first layer of a neural network is called the
\textit{input} layer, the intermediate layers are called the
\textit{hidden} layers and the last layer is called the
\textit{output} layer.

In order to learn the relationship between $(X,Y)$, learnable weights
are used for each neuron. If a single neuron has inputs $x_i$, then
each $x_i$ has a weight $w_i$ associated to this input. Additionally,
a single, learnable \textit{bias} parameter is added in order to
increase the flexibility of the model to fit the data. This process in
a neuron, consisting of the weights $w_i$ and the bias $b$, shows a
linear response:
\begin{equation}
f_\Sigma = \sum_{i=1}^{n}w_i x_i + b.
\end{equation}
However, many physical processes in nature are inherently
nonlinear. To account for this, the output of the transfer function can be
wrapped in another, nonlinear function. Additionally, it can be shown
that a nonlinear, two-layer neural network can approximate any
function (Chap.~6.4.1. in Ref.~\autocite{Goodfellow2016}).
The most commonly used, nonlinear function is the \textit{rectified
  linear unit} (ReLU):
\begin{equation}
f_{\text{ReLU}}(x) = \max (0,x).
\end{equation}
Such functions are called \textit{activation} functions or layers.

The weights of each neuron get updated iteratively during a training
process. For this purpose, one needs to define a so-called \textit{cost} or
\textit{loss function}, which measures the distance between the output of the
neural network $f(x) = y_{\text{reco}}$ and the ground truth
$y_{\text{true}}$. This can for example be done by measuring the mean squared
error and minimising $(y_{\text{true}}-y_{\text{reco}})^2$.

Typically, iterative gradient descent (Chap.~4.3. in
Ref.~\autocite{Goodfellow2016}) based optimisation algorithms are used
that minimise the cost function until a low value is achieved. During
this training process, the cost error is back-propagated using a
back-propagation algorithm (Chap.~6.5. in
Ref.~\autocite{Goodfellow2016}), which allows for the tuning of the
neural network's weights.

Convolutional neural networks are frequently used in domains, where the
input can be expected to be image-like, i.e. in image or video
classification.  Therefore, several changes are made in the
architecture of CNNs compared to fully connected neural networks. The
main concepts of convolutional neural networks are based on only
locally and not fully connected networks and on parameter sharing
between certain neurons in the network.

Typical input images to convolutional neural networks are
two-dimensional (2D). However, since most images are coloured, they in
fact are encoded by three dimensions (3D): width, height and
channel. Here, the channel dimension specifies the brightness for each
colour channel (red, green and blue) of the image. This three
dimensional array is then used as input to the first convolutional
layer.

Similar to the input layer, the neurons in a convolutional layer are
also arranged in three dimensions, called width, height and depth.  As
already mentioned, one of the main differences between convolutional
layers and fully connected layers is that the neurons inside a
convolutional layer are only connected to a local region of the input
volume. This is often called the \textit{receptive field} of the
neuron.  The connections of this local area to the neuron are local in
space (width, height), but they are always full along the depth of the
input volume. Hence, for a $\text{[32}\times\text{32}\times\text{3]}$
image (width, height, channel) and receptive field size of 5 pixels,
each neuron in the first convolutional layer is connected to a local
$\text{[5}\times\text{5}\times\text{3]}$ (width, height, depth)
patch of the input. To each of these connections, a weight is assigned, such
that each neuron has a [$\text{5} \times \text{5} \times \text{3}$]
weight matrix, which is often called the \textit{kernel} or
\textit{filter}. These weights are used in performing a dot product
between the receptive field of the neuron and its associated kernel,
also called the \textit{convolution} process. Here, the total number
of parameters for the single neuron would be $\text{5} \cdot \text{5}
\cdot \text{3} + \text{1} ~(\text{bias}) = \text{76}$. Additionally,
each neuron at the same depth level covers a different part of the
image with its receptive field. For more information the reader is
referred to Refs.~\autocite{Goodfellow2016, cs231n}.

For CNNs, an important assumption is that abstract image structures,
such as edges, occur multiple times in the image.  Under this
assumption, the neurons at a certain depth can share their weights,
which significantly reduces the number of parameters in the network.
The number of parameters in CNNs can be further reduced with the aid
of \textit{pooling} layers, which reduce the dimensionality of the
layer outputs by selecting or combining the neuron outputs
\autocite{Goodfellow2016, cs231n}.

For the training of a neural network, the data are split up into a 
training, validation, and test dataset. The network is trained on the
training dataset and validated by applying the network to
the validation dataset. Once the training is finished, the network is
applied to the test dataset to determine its performance. If the value
of the loss function is significantly greater for the training dataset
than for the validation dataset, this is called \textit{overfitting}
with respect to the training dataset. The smaller the size of the
training dataset, the higher the probability that the network will
focus on the peculiarities of individual input images instead of
generalising generic image features. In order to avoid overfitting,
so-called regularisation techniques such as \textit{dropout} layers
have been developed \autocite{JMLR:v15:srivastava14a}. In a dropout
layer, inputs are randomly set to zero with a probability defined by
the \textit{dropout rate} $\delta$.
In order to set up a basic convolutional neural network, the
convolutional layers are stacked.  After the last convolutional layer,
the multi-dimensional output array is reshaped without ordering into a
one-dimensional array by a \textit{flattening layer}. A small
fully-connected network can then be added, in order to connect the
outputs of the last convolutional layer to the output neurons of the
full network.

\section{Data pre-processing}

For each hit in a simulated event, the PMT identifier, i.e. the
relative coordinate of the hit PMT in a DOM, is recorded.
Additionally, the time at which the PMT signal crosses the discriminator
threshold, and the measured ToT, are stored. However, the ToT value
itself is not used as input for the CNNs.  In order to feed this
four-dimensional event data to a CNN, the hits can be binned into
rectangular pixels, such that each image encodes three spatial
dimensions (XYZ) and the time dimension T.

\subsection{Spatial binning}
The number of pixels required to resolve the spatial coordinates of
the individual PMTs inside the DOMs would be very large, and most bins
would be empty due to the sparsely instrumented detector volume.
Therefore, the pixelation is defined such that exactly one DOM fits
into one bin, while some bins remain empty due to the detector
geometry. In the case of the full KM3NeT/ORCA detector, this results
in a $\text{11} \times \text{13} \times \text{18}$ (XYZ) pixel grid.
The information regarding which PMT in a DOM has been hit is
discarded. This is corrected for by adding a PMT identifier dimension to the
pixel grid, resulting in a XYZP grid. Since one DOM holds 31 PMTs, the
final spatial shape of such an image is $\text{11} \times \text{13}
\times \text{18} \times \text{31}$ (XYZP). Such image types can also
be found in classical computer vision tasks, e.g. as coloured
videos. The only difference is that in a conventional video the
Z-coordinate is replaced by the time and the PMT identifier is
replaced by the red-green-blue colour information.

\subsection{Temporal Binning}
\label{subsect:temporal_binning}

The indispensable piece of information still missing in these
images is the time at which a hit has been recorded. This
information can be added as an additional dimension, such that the
final image of an event is five-dimensional: XYZTP.

The time resolution in KM3NeT/ORCA is of the order of nanoseconds
\autocite{Aiello:2018nvl}. As explained in
section~\ref{sect:simulations_and_trigger}, the time length of an
event is about \SI{3}{\micro\second}, implying 3000 bins for the time
dimension to reach nanosecond resolution.  However, with each
additional bin, the size of the event image gets larger and this leads
to additional computations in the first layer of a CNN. Hence, the
number of bins of the final image should be as low as possible, while
still containing the relevant timing information.

\begin{figure}[h!]
	\centering \centering
        \includegraphics[width=0.675\textwidth]{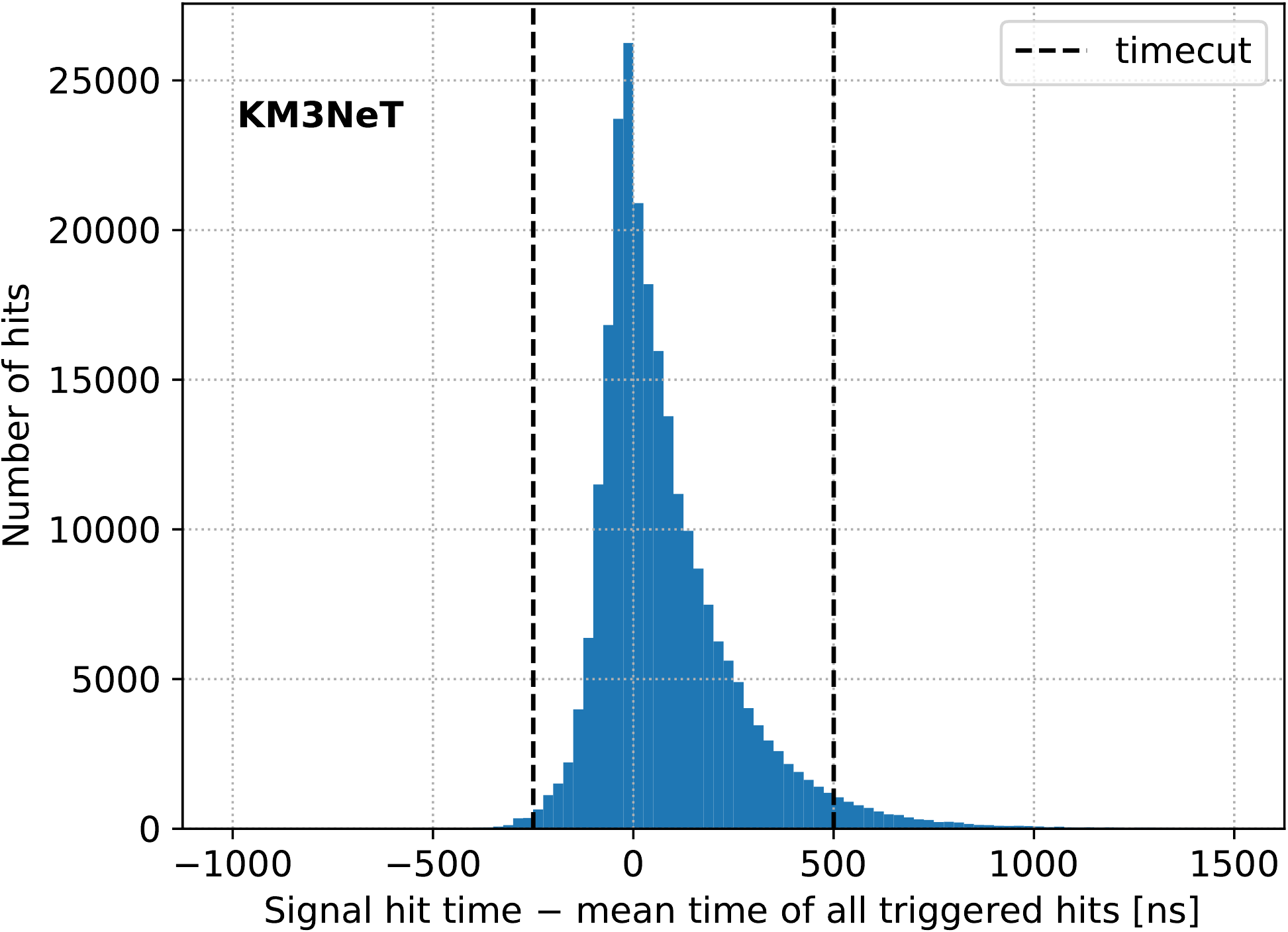}
	\caption[Time distribution of muon neutrino signal hits
          centered with the mean time of the triggered hits for single
          events.]{Time distribution of $\numuCC$ signal hits relative
          to the mean time of the triggered hits calculated for each
          individual event. For this distribution, about 3000
          $\numuCC$ events in the energy range from
          \SIrange{3}{100}{\giga\electronvolt} have been used,
          cf.~\mytref{tab:TabSim1}. The dashed line in black indicates
          a possible timecut for the time range used to generate the
          CNN input images.}
	\label{fig:timecut_muon-CC}
\end{figure}

Most hits that lie outside the time range of the triggered hits are
background hits and not signal hits.
%
Therefore, background hits can be discriminated against
to some extent by selecting the time range in which most signal hits
are found. Investigating the distribution of the time of the signal
hits relative to the mean of the triggered hit times in individual
events, as depicted in \myfref{fig:timecut_muon-CC} for $\numuCC$
events, shows that it is asymmetric and that the relevant time range
can be reduced significantly for the image generation binning.

Since the time range covered by the triggered hits is different for
each event, the time range selection is defined relative to the mean
time of the triggered hits for each event.  As can be seen in
\myfref{fig:timecut_muon-CC}, only a small fraction of the signal hits
are removed, as indicated by the timecut defined by the dashed, black
lines. A compromise needs to be found between the width of the
timecut window and the number of time bins, implying a certain time resolution
available to the network. The specific values of the timecuts used in
this work are reported in the respective image generation sections of
the presented CNNs.  The timecut window is a parameter that can
be further optimised with respect to the final performance of a
trained neural network.
In this
work, no such parameter optimisation studies have been carried out for
any of the presented CNNs, which implies that their performance for
specific use cases can likely be improved further.

\subsection{Multi-image Convolutional Neural Networks}
\label{sec:preprocessing:multi}
After binning, the resulting images are five-dimensional: XYZTP.  In
order to train the neural networks presented in this paper, the deep
learning framework TensorFlow \autocite{tensorflow2015-whitepaper} has
been used in conjunction with the Keras \autocite{chollet2015keras}
high-level neural network programming library.  However, TensorFlow
does not support convolutional layers
which accept more than four dimensions as input,
since five dimensional inputs are not a
usual case in computer vision.
Hence, the five dimensions of the XYZTP
images need to be reduced to four dimensions, such that one can use
three-dimensional convolutional layers. To this end, one image
dimension is summed up, i.e.  the information of individual PMTs in a
DOM is discarded, such that the resulting image is only
four-dimensional (XYZT). However, a second image of the same event is
then fed to the network (XYZP) that recovers the information regarding
which PMT in a DOM has been hit, but discards its hit time.  Since
these images only differ in the 4th dimension, i.e. the depth
dimension of a convolutional layer, the images can be stacked in this
dimension. For example, an XYZP image of dimension $\text{11} \times
\text{13} \times \text{18} \times \text{31}$ can be combined with an
XYZT image of dimension $\text{11} \times \text{13} \times \text{18}
\times \text{N}_\text{T}$ into a single, stacked XYZ-T/P image of
dimension $\text{11} \times \text{13} \times \text{18} \times
(\text{N}_\text{T}+\text{31})$.
These images lack the information about the hit time for a specific PMT,
if more than one hit has occurred on a DOM in an event.

Significant gains in performance for all CNN applications
in this work were observed when using this stacking method,
as compared to just supplying a single XYZT image. Furthermore, it
will be demonstrated that networks with such input limitations can
still match or outperform the KM3NeT/ORCA reconstruction algorithms as
presented in Ref.~\autocite{Adrian-Martinez2016}.

\section{Main network architecture} 
\label{sect:main_cnn_arch}

Four-dimensional images that have been created from simulated events
are fed as input to a CNN.  All networks that have been designed for a
specific task in this work share a common architecture.  The CNNs
consist of two main components: the convolutional part with the
convolutional layers and a small fully-connected network in the end.
The convolutional part consists of \textit{convolutional blocks}, each
of them containing a convolutional layer, a batch normalisation layer
\autocite{Ioffe2015}, an activation layer, and, optionally, a dropout
\autocite{JMLR:v15:srivastava14a} or a pooling layer.  The batch
normalisation layer usually enables a faster and more robust training
process of deep neural networks by normalising, scaling and shifting
the output of the convolutional layer. The scaling and shifting
transform is controlled by learnable parameters during the training
process. Recent studies indicate that the batch normalisation method
smoothes the optimisation landscape and induces a more predictive and
stable behaviour of the gradients, allowing for faster training
\autocite{2018arXiv180511604S}.

In the three-dimensional convolutional layer, the weights are
initialised based on a uniform distribution whose variance is
calculated according to Ref.~\autocite{He2015}, while the biases are
set to zero.
Additionally, the kernel size is three ($\text{3} \times
\text{3} \times \text{3}$), the stride, i.e. the step size in shifting
the convolutional kernel, is one ($\text{1} \times \text{1} \times
\text{1}$) and zero-padding (Chap.~9 in
Ref.~\autocite{Goodfellow2016}) is used.  For the batch normalisation
layers, the standard parameters from Ref.~\autocite{Ioffe2015} are
used. After this, a ReLU activation layer is added. These three layers
are found in all convolutional blocks that are used in this
work. Additionally, optional maximum pooling and dropout layers are
added. In the case of maximum pooling layers, zero-padding is not
applied.  A scheme of these convolutional blocks is shown in
\mytref{tab:conv_block_scheme_1}.

\setlength{\dashlinedash}{0.2pt}
\setlength{\dashlinegap}{4.5pt}
\setlength{\arrayrulewidth}{0.2pt}
\renewcommand{\arraystretch}{1.1}
\vspace{0.3cm}
\begin{table}[h!]
	\centering
	\setlength{\extrarowheight}{0.1cm}
	\begin{tabular}{|c|c|}
		\hline 		
		\rowcolor{Gray} Layer type & Properties \\ 
		\hline
		Convolution & kernel size ($\text{3} \times \text{3} \times \text{3}$),
                uniform initialisation \autocite{He2015}, zero-padding   \\ \hdashline
		Batch normalisation & parameters as in Ref.~\autocite{Ioffe2015} \\ \hdashline
		Activation & ReLU \\ \hdashline
		Maximum pooling & optional, no zero-padding  \\ \hdashline
		Dropout & optional \\ \hline
		
	\end{tabular}
        \vspace{0.3cm}
	\caption[Structure of a convolutional block.]{Scheme of a
          convolutional block used for all CNNs defined in this work.}
	\label{tab:conv_block_scheme_1}
\end{table}
\renewcommand{\arraystretch}{1.0}

\noindent Furthermore, all models use the Adam gradient descent optimiser
\autocite{Kingma2014} with standard parameter values, except for the
parameter $\epsilon$, which is increased from its default value of
$\text{10}^{\text{-8}}$ to $\text{10}^{\text{-1}}$. A larger value of
$\epsilon$ results in smaller weight updates after each training step.
In our case, it has been observed that the network occasionally did
not start to learn, depending on the random initialisation of the
parameters.  This could be fixed by changing the value of $\epsilon$
to $\text{10}^{\text{-1}}$ as suggested in Ref.~\autocite{epsilon},
while significant drawbacks, such as a slower training convergence due
to smaller weight updates, have not been observed.  The weights of the
neural network are updated after one \textit{batch} of images is
passed through the network. This is known as batch gradient descent.
The \textit{batch size} is defined as the number of images contained
in one batch.  The batch size in the training for all presented CNNs
is generally set to 64 and the learning rate, i.e. the step size in
the Adam algorithm for the update of the weights
is annealed exponentially.

The training of all presented networks has been executed at the
TinyGPU cluster at the RRZE computing centre \footnote{Regionales
  Rechenzentrum Erlangen}. It consists of 32 nodes with 4 GPUs each.
The GPUs are either Nvidia GTX1080, GTX1080Ti, RTX2080 Ti, or Tesla
V100.  All CNNs in this work have been trained with CUDA~10
\autocite{Cuda}.  In order to train the networks, an open-source
software framework called \textit{OrcaNet} \autocite{orcanet} has been
developed, which is intended as a high-level application programming
interface on top of Keras \autocite{chollet2015keras}, specifically
suited to the large datasets that frequently occur in astroparticle
physics.

\section{Background classifier}
An essential part of the KM3NeT/ORCA reconstruction pipeline is the
background classifier, which discriminates atmospheric muons and
random noise from neutrino-induced events. For this purpose, the
employed classification algorithm is based on a Random
Forest (RF) \autocite{Breiman2001} method. The inputs of the RF are
high-level observables (features), mainly determined from
likelihood-based track and shower reconstruction algorithms. In this
section, an alternative classifier based on CNNs is presented and its
performance is compared to the RF classifier.

%
\subsection{Image generation}
\label{subsect:bg_image_gen}

\begin{figure}[h!]
	\centering
        \includegraphics[width=0.675\textwidth]{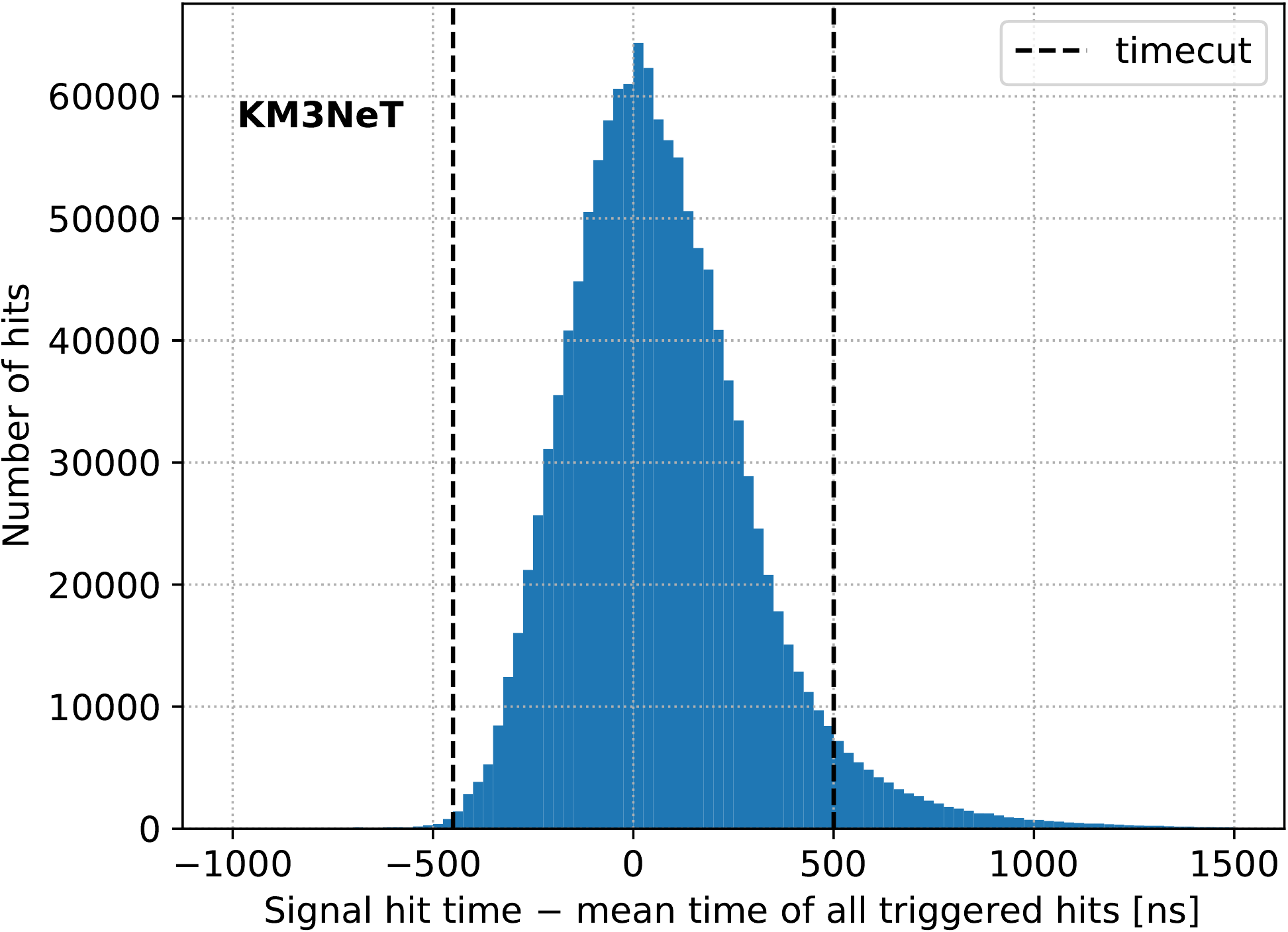}
	\caption[Time distribution of atmospheric muon signal hits
          centered on the mean time of the triggered hits for single
          events.]{Time distribution of signal hits in atmospheric
          muon events relative to the mean time of the triggered hits
          calculated for each individual event. For this distribution,
          about 3000 atmospheric muon events have been used. The
          dashed black line indicates the timecut, set to an interval
          of [$-\SI{450}{\nano\second}, + \SI{500}{\nano\second}$],
          that has been applied for the generation of the background
          classifier images.}
	\label{fig:timecut_mupage}
\end{figure}

As outlined in \mysref{sec:preprocessing:multi},
XYZ-T/P images
are used as input to the network.
For both event images, i.e. the XYZT and XYZP components of the
stacked XYZ-T/P image, a timecut has been defined, as introduced in
\mysref{subsect:temporal_binning}. The signal hit time distribution of
atmospheric muons, relative to the mean time of all triggered hits, is
shown in \myfref{fig:timecut_mupage}.  This distribution has a larger
variance than for neutrino events shown in
\myfref{fig:timecut_muon-CC}. The reason is that, on average,
atmospheric muons traverse larger parts of the detector compared to
the secondary particles of the GeV-scale neutrino interactions of
interest. Hence, the timecut window for all event classes has been
set conservatively based on atmospheric muon events, resulting in a
width of \SI{950}{\nano\second}.

The timecut for this distribution, as indicated in
\myfref{fig:timecut_mupage}, has been set to an interval of
       [$-\SI{450}{\nano\second}, + \SI{500}{\nano\second}$], keeping
       more early signal hits than late hits. The reason for this is
       that events which produce late hits are typically energetic and
       leave a longer trace in the detector. Hence, they can be better
       reconstructed than less energetic atmospheric muons that only
       produce a few hits at the edge of the detector. Consequently,
       cutting away a few late hits has a small effect compared to
       discarding early hits from a low-energy atmospheric muon event,
       which already produces a low number of hits.

The number of time bins is set to 100, such that the XYZT
images have dimensions of $\text{11} \times \text{13} \times \text{18}
\times \text{100}$. The resulting time resolution of each time
bin is \SI{9.5}{\nano\second}, which translates to about
\SI{2}{\metre} of photon propagation distance.
Adding the information of
the 31 PMTs per DOM, as described in \mysref{sec:preprocessing:multi},
yields final event images that have a dimension of $\text{11} \times
\text{13} \times \text{18} \times \text{131}$.

\subsection{Network architecture}
\label{subsect:bg_network_arch}

The CNN network architecture for the background classifier is based on
the three-dimensional convolutional blocks introduced in
\mysref{sect:main_cnn_arch}, with two additional fully-connected layers, also called \textit{dense} layers, at the end. The output layer of the CNN is composed of
two neurons, such that the network only distinguishes between neutrino
and non-neutrino events.
An overview of the final network structure is shown in \mytref{tab:bg_model_xyztc_1}.

\renewcommand{\arraystretch}{1.1}
\begin{table}[h!]
	\centering
	\setlength{\extrarowheight}{0.1cm}
	\setlength{\tabcolsep}{0.4cm}
	\begin{tabular}{|c|rcl|}
		
		\multicolumn{1}{c}{Building block / layer} & \multicolumn{3}{c}{Output dimension} \\ \hline
		\rowcolor{Gray} XYZ-T Input & 11 $\times$ 13 $\times$ 18 & $\times$ & 100 \\ 
		\rowcolor{Gray} XYZ-P Input & 11 $\times$ 13 $\times$ 18 & $\times$ & 31 \\ 
		\rowcolor{Gray} Final stacked XYZ-T + XYZ-P Input & 11 $\times$ 13 $\times$ 18 & $\times$ & 131 \\ \hline
		
		Convolutional block 1 (64 filters) & 11 $\times$ 13 $\times$ 18 & $\times$ & 64 \\ \hdashline
		Convolutional block 2 (64 filters) & 11 $\times$ 13 $\times$ 18 & $\times$ & 64 \\ \hdashline
		Convolutional block 3 (64 filters) & 11 $\times$ 13 $\times$ 18 & $\times$ & 64 \\ \hdashline
		Convolutional block 4 (64 filters) & 11 $\times$ 13 $\times$ 18 & $\times$ & 64 \\ \hdashline
		Convolutional block 5 (64 filters) & 11 $\times$ 13 $\times$ 18 & $\times$ & 64 \\ \hdashline
		Convolutional block 6 (64 filters) & 11 $\times$ 13 $\times$ 18 & $\times$ & 64 \\ \hdashline
		Max pooling (2,2,2) & 5 $\times$ 6 $\times$ 9 & $\times$ & 64 \\ \hline
		
		Convolutional block 1 (128 filters) & 5 $\times$ 6 $\times$ 9 & $\times$ & 128 \\ \hdashline 
		Convolutional block 2 (128 filters) & 5 $\times$ 6 $\times$ 9 & $\times$ & 128 \\ \hdashline 
		Convolutional block 3 (128 filters) & 5 $\times$ 6 $\times$ 9 & $\times$ & 128 \\ \hdashline 
		Convolutional block 4 (128 filters) & 5 $\times$ 6 $\times$ 9 & $\times$ & 128 \\ \hdashline 
		Max pooling (2,2,2) &  2 $\times$ 3 $\times$ 4 & $\times$ & 128 \\ \hline
		
		Flatten & \multicolumn{3}{c|}{3072}  \\ \hdashline 
		Dense  + ReLU & \multicolumn{3}{c|}{128} \\ \hdashline 
		Dense  + ReLU \quad & \multicolumn{3}{c|}{32} \\ \hdashline 
		Dense  + Softmax & \multicolumn{3}{c|}{2} \\ \hline
		
	\end{tabular}
	\vspace{2mm}
	\caption{Network structure of the background classifier's
          three-dimensional CNN model with XYZ-T/P input. No dropout
          is used due to the large training dataset of
          $\text{42.6}\times\text{10}^{\text{6}}$ training events.}
	\label{tab:bg_model_xyztc_1}
\end{table}
\renewcommand{\arraystretch}{1.0}

Initially, a CNN with three output neurons was tested, so that
neutrinos, atmospheric muons and random noise events could be
classified separately. However, it was observed that the three-class
CNN performed slightly worse than the two-class CNN that distinguishes
only neutrino events from all others. This is due to the fact that the
network cannot prioritise neutrino versus non-neutrino classification in
the three-class case. A mistakenly classified atmospheric muon, e.g.,
classified as a random noise event, has the same effect on the total
loss as an atmospheric muon classified as a neutrino.

No regularisation techniques, such as dropout, are added to
the network.
The training dataset is large enough,
cf.~\mysref{subsect:dsplit_bg}, so that virtually no overfitting
occurs, i.e. the training-phase loss is of the same order as the loss
during the validation phase.

\subsection{Preparation of training, validation and test data}
\label{subsect:dsplit_bg}

For the training of the background classifier, the simulated data from
\mytref{tab:TabSim1}
is split into a training, validation and test dataset.

In order to
balance the datasets with respect to their class frequency, one could
split the data into 50\% neutrino and 50\% non-neutrino events (25\%
atmospheric muons + 25\% random noise).  Considering that the neutrino
sample has the lowest number of events (about
$\text{22.8}\times\text{10}^{\text{6}}$), one would have to remove a
significant fraction of the $\text{23.3}\times\text{10}^{\text{6}}$
generated random noise events, and of the
$\text{65.1}\times\text{10}^{\text{6}}$ atmospheric muon events for a
class-balanced data splitting.  On the other hand, based on the RF
background classifier, it can be expected that the final accuracy of
the classifier should be close to 99\%. Therefore, a balanced
splitting of the data into 50\% for each class is not necessary, in
order to avoid a local minimum during the training process.
The following data splitting is used: 1/3 neutrino events, 1/3 random
noise events and 1/3 atmospheric muon events, and hence the final
class balance is 1/3 neutrino events and 2/3 non-neutrino events.
Using this data splitting and considering the number of MC events
summarised in \mytref{tab:TabSim1}, the size of the used training
dataset is larger compared to a 50/50 split. The fractions of
different neutrino flavours and interaction types is kept as indicated
in \mytref{tab:TabSim1}.

This rebalanced dataset is then split into 75\% training, 2.5\%
validation, and 22.5\% test events, which is a trade-off between
maximizing the training dataset and retaining sufficient statistics
for performance evaluations.  Additionally, the events that have been
removed to balance the dataset (mostly atmospheric muons) are added to
the test dataset.  In total, the training data contains about
$\text{42.6}\times\text{10}^{\text{6}}$ events.

Using a Nvidia Tesla V100 GPU, it takes about a week to fully train
this CNN background classifier. The time needed for the training
scales more weakly than linear with the number of time bins, which 
can be increased to improve the time resolution of the input images.

\subsection{Performance and comparison to Random Forest classifier}
\label{subsect:bg_performance}

\begin{figure}[h!]
\centering
\begin{overpic}[width=0.675\linewidth,page=1]{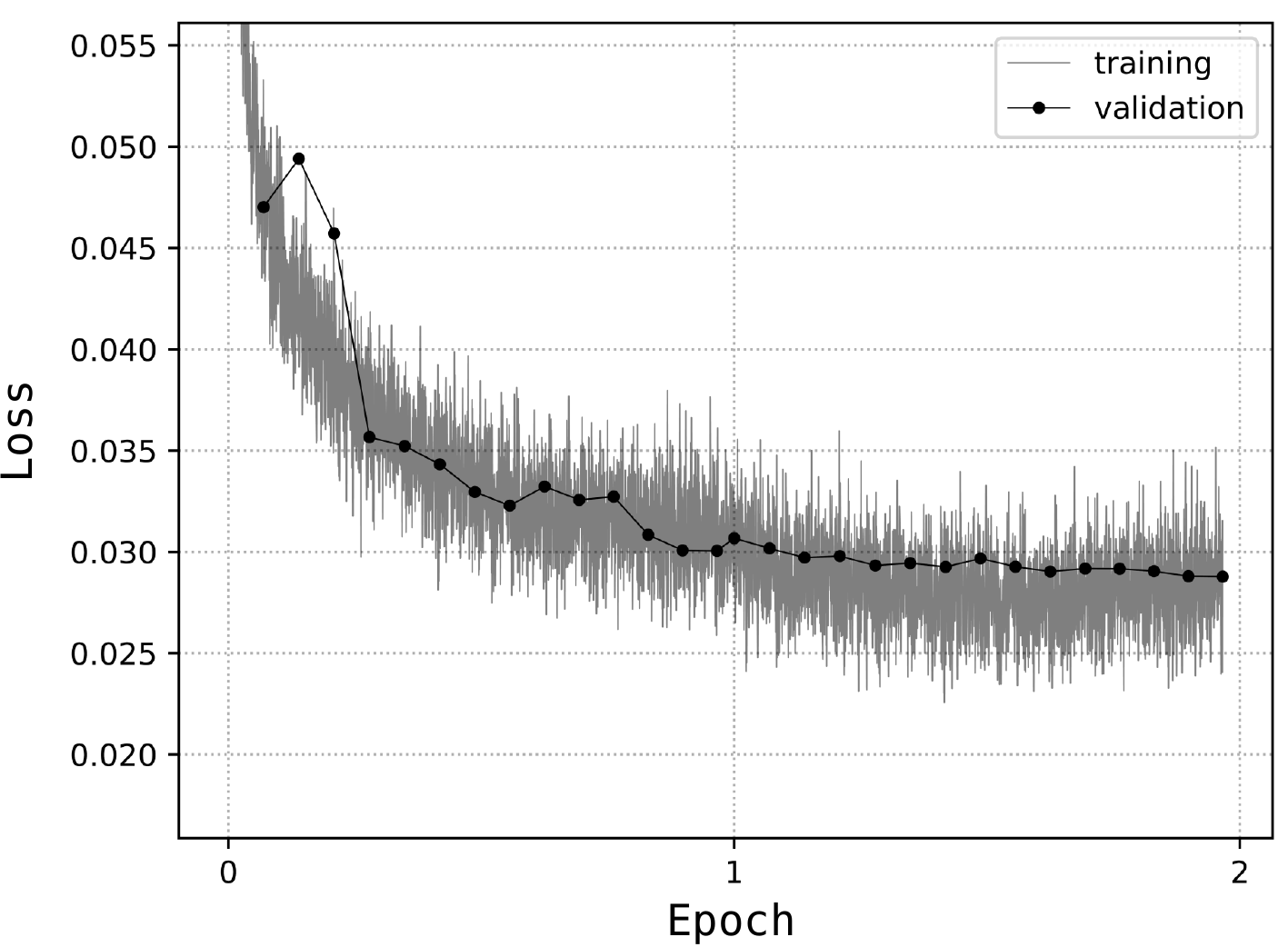}
\put (40,66) {\textsf{KM3NeT}}
\end{overpic}
\caption{Training and validation cross-entropy loss of the
          background classifier during the training.  Each data point
          of the training loss curve is averaged over 250 batches,
          i.e. $\text{250}\times\text{64}$ event images.}
\label{fig:bg_loss}
\end{figure}

The performance of the CNN background classifier is evaluated using
the training and validation cross-entropy loss
\autocite{Goodfellow2016} of a specific classifier, the
\textit{softmax classifier} \autocite{Goodfellow2016}, as a function
of the number of epochs, and is shown in \myfref{fig:bg_loss}. An
epoch is defined as one training process of the CNN, using the entire
training event dataset.  The training is stopped after approximately
two epochs, as the validation loss shows no further significant
improvement. At the end of the training no overfitting is observed.

\noindent In order to compare the CNN performance to the RF background
classifier, the same test dataset is used.
%
A pre-selection of these events is carried out to reduce the fraction of
atmospheric muon and random noise events to a few percent of all
triggered events. For atmospheric muons this is achieved by selecting
only events for which the reconstructed particle direction is below
the horizon, i.e. which are \textit{up-going} events in the detector.
Furthermore, the events must have been reconstructed with high quality
by the KM3NeT/ORCA maximum-likelihood reconstruction algorithms for
either track-like or shower-like events.  Finally, events
reconstructed by the maximum-likelihood-based algorithms as
originating from outside of the instrumented volume of the detector
are removed.

In total, the pre-selected test dataset consists of about
$\text{3.3}\times \text{10}^{\text{6}}$
neutrino, about $\text{6}\times \text{10}^{\text{4}}$
atmospheric muon and about $\text{4}\times \text{10}^{\text{4}}$
random noise events.
%
%
This selection is used for
all of the following performance evaluations.

\begin{figure}[h!]
	\centering \includegraphics[width=0.675\textwidth,
          page=1]{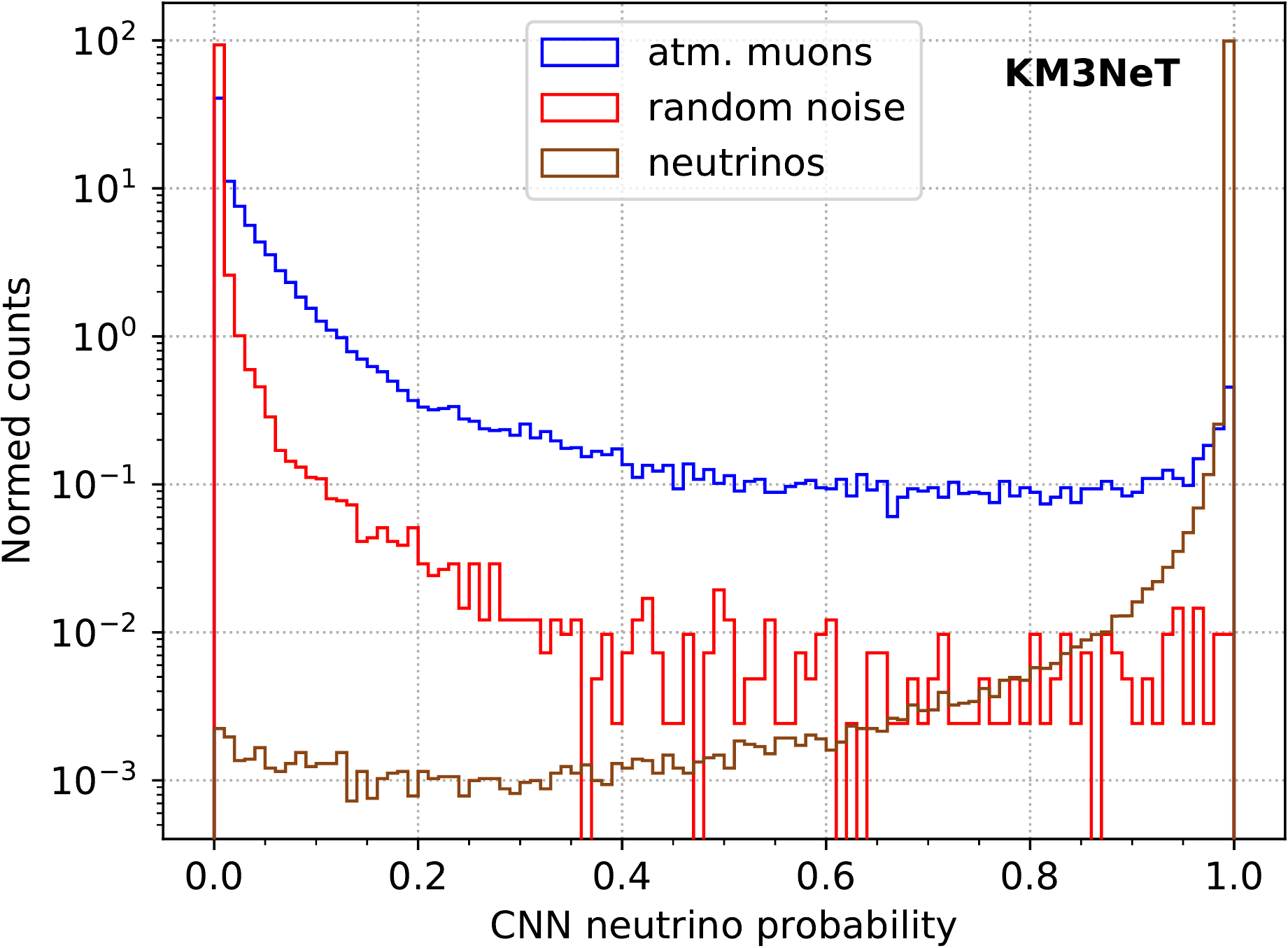}
	\caption{Distribution of the CNN neutrino probability for
          pre-selected atmospheric muon (blue), random noise (red),
          and neutrino (brown) events. All three distributions have
          been normalised to the area under each histogram.}
	\label{fig:prob_neutrino}
\end{figure}

In order to get a first impression of the CNN-based background
classifier, the distribution of the neutrino class probability is
investigated for all three event classes (neutrinos, atmospheric muons,
random noise), cf.~\myfref{fig:prob_neutrino}.

Based on the results shown in \myfref{fig:prob_neutrino}, it can be
seen that the rate of random noise events misclassified as
neutrino-induced events is significantly lower than for atmospheric
muons.  Using the predicted probability for an event to be classified
as neutrino-induced, a threshold value $p$ can be set to remove
background events.

In order to quantify the performance of the CNN background classifier,
the metric shown below is used to investigate the fraction of remaining
atmospheric muon and random noise events for a given threshold value
$p$.  The atmospheric muon or random noise contamination, and the
neutrino efficiency, are defined as:

\begin{equation}
\text{C}_{\upmu / \text{RN}}(p) = \frac{N_{\upmu/\text{RN}}(p)}{N_{\text{total}}(p)}\;,
\end{equation}
\begin{equation}
\upnu_{\text{eff}}(p) = \frac{N_{\upnu}(p)}{N_{\upnu, \text{total}}}\;.
\end{equation}

Here, $N_{\upmu / \text{RN}}$ is the number of atmospheric muon or
random noise events, whose probability to be a neutrino-induced event
is higher than $p$, while $N_{\text{total}}(p)$ accounts for the total
number of events, after the same cut on $p$.  Regarding the neutrino
efficiency, $N_{\upnu}(p)$ is the total number of neutrinos in the
dataset whose neutrino probability is greater than the threshold value
$p$, while $N_{\upnu, \text{total}}$ is the number of neutrinos in the
dataset, without applying any threshold.

\begin{figure}[h!]
	\centering \includegraphics[width=0.675\textwidth]{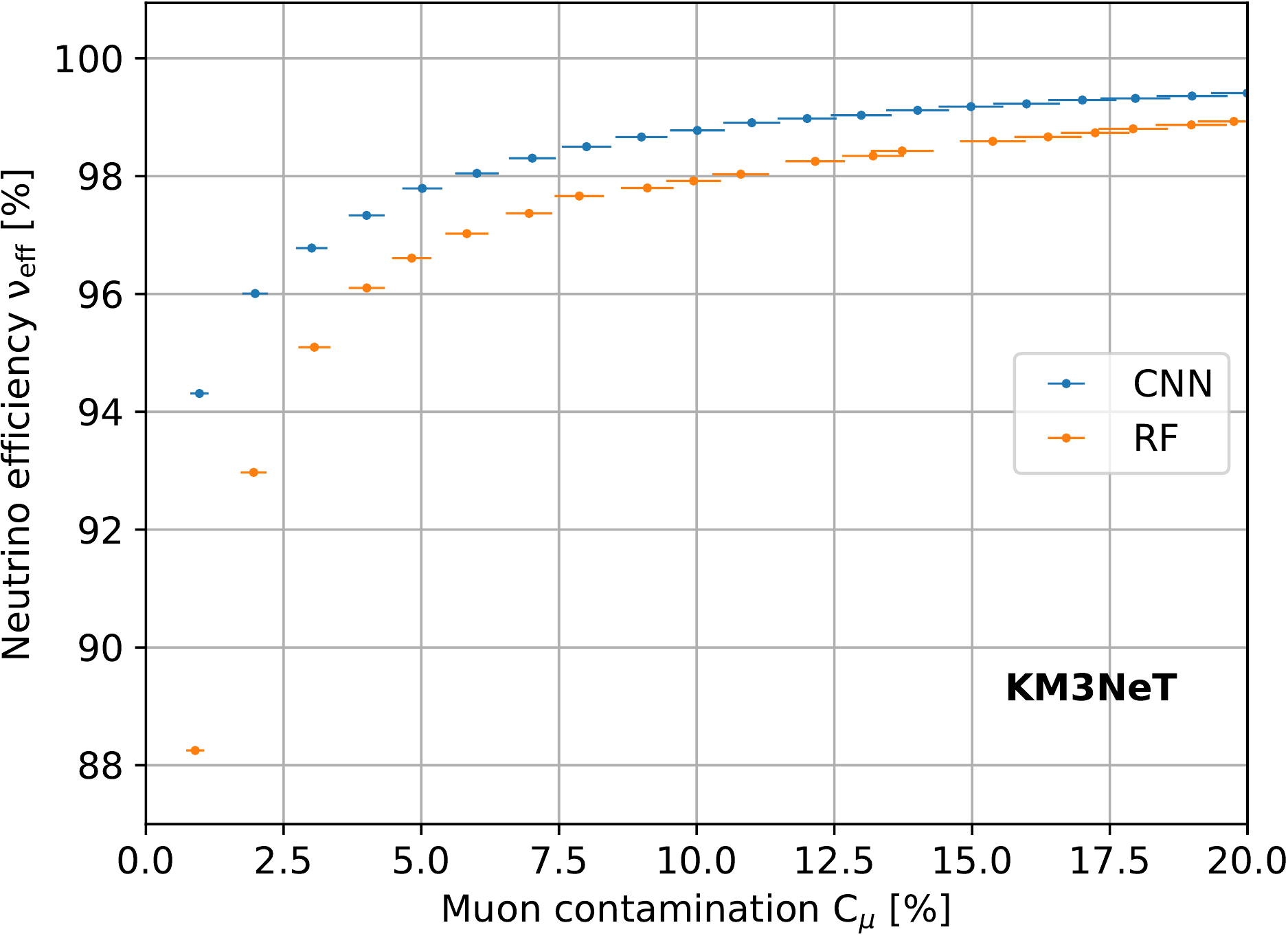}
	\caption{Neutrino efficiency, $\upnu_{\text{eff}}$, versus
          atmospheric muon event contamination, $\text{C}_{\upmu}$,
          weighted with the Honda atmospheric neutrino flux
          \autocite{HKKM} model. The CNN (RF) performance is depicted
          in blue (orange). All neutrinos from the pre-selected test
          dataset are included
          (\SIrange{1}{100}{\giga\electronvolt}).
        }
	\label{fig:muon_cont_vs_neutr_eff}
\end{figure}

Based on the results shown in \myfref{fig:muon_cont_vs_neutr_eff} for
the neutrino efficiency $\upnu_{\text{eff}}$ as a function of the
residual atmospheric muon contamination $\text{C}_{\upmu}$, it can be
concluded that the CNN background classifier yields a higher neutrino
efficiency, of the order of a few percent, for the same muon
contamination compared to the RF background classifier.

\begin{figure}[h!]
	\centering \includegraphics[width=0.675\textwidth]{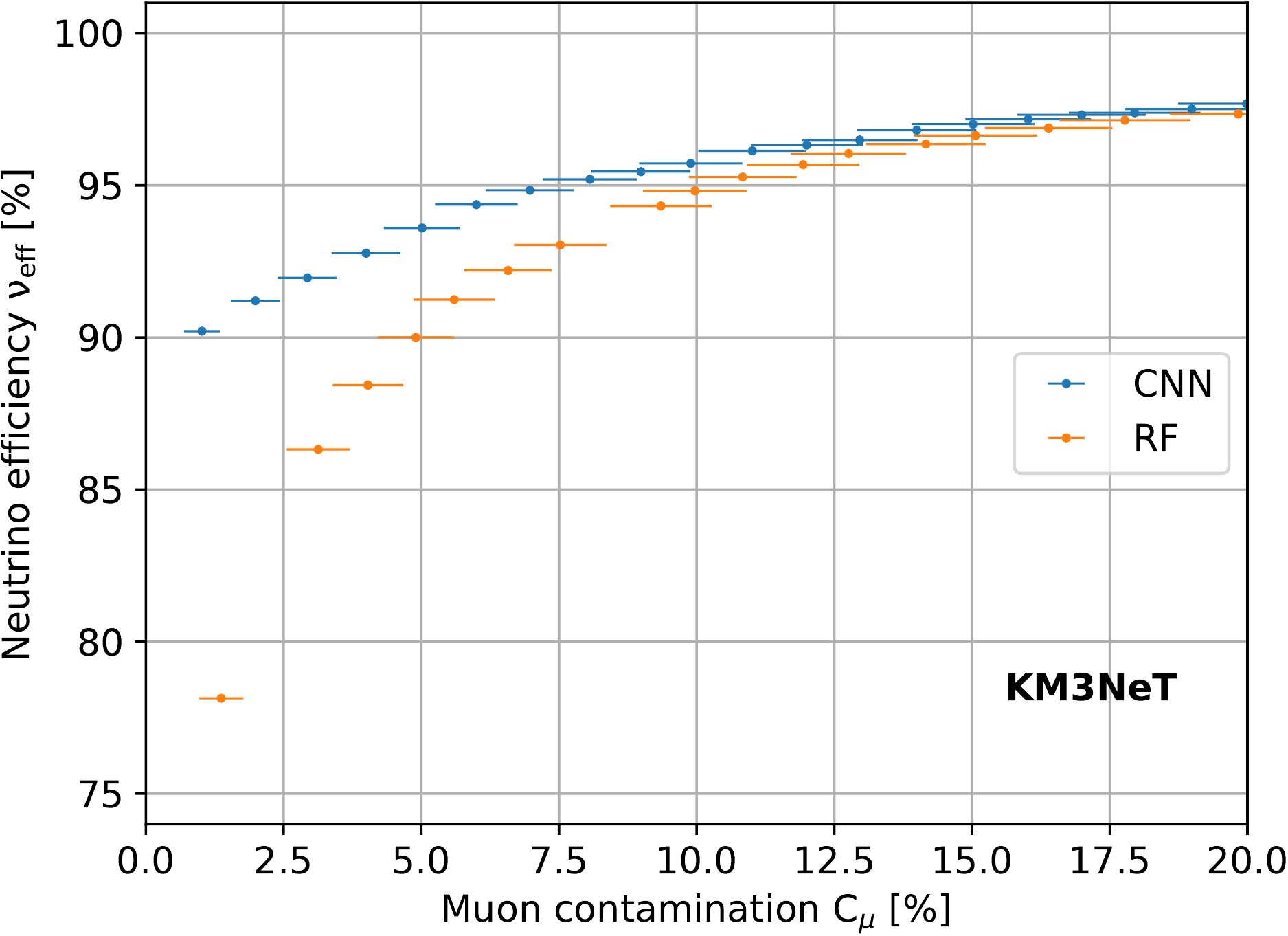}
	\caption{Neutrino efficiency, $\upnu_{\text{eff}}$, versus
          atmospheric muon event contamination, $\text{C}_{\upmu}$,
          weighted with the Honda atmospheric neutrino flux
          \autocite{HKKM} model. The CNN (RF) performance is depicted
          in blue (orange). Only neutrinos with a MC energy in the
          range of \SIrange{1}{5}{\giga\electronvolt} have been used.}
	\label{fig:muon_cont_vs_neutr_eff_1_5}
\end{figure}

\begin{figure}[h!]
	\centering
        \includegraphics[width=0.675\textwidth]{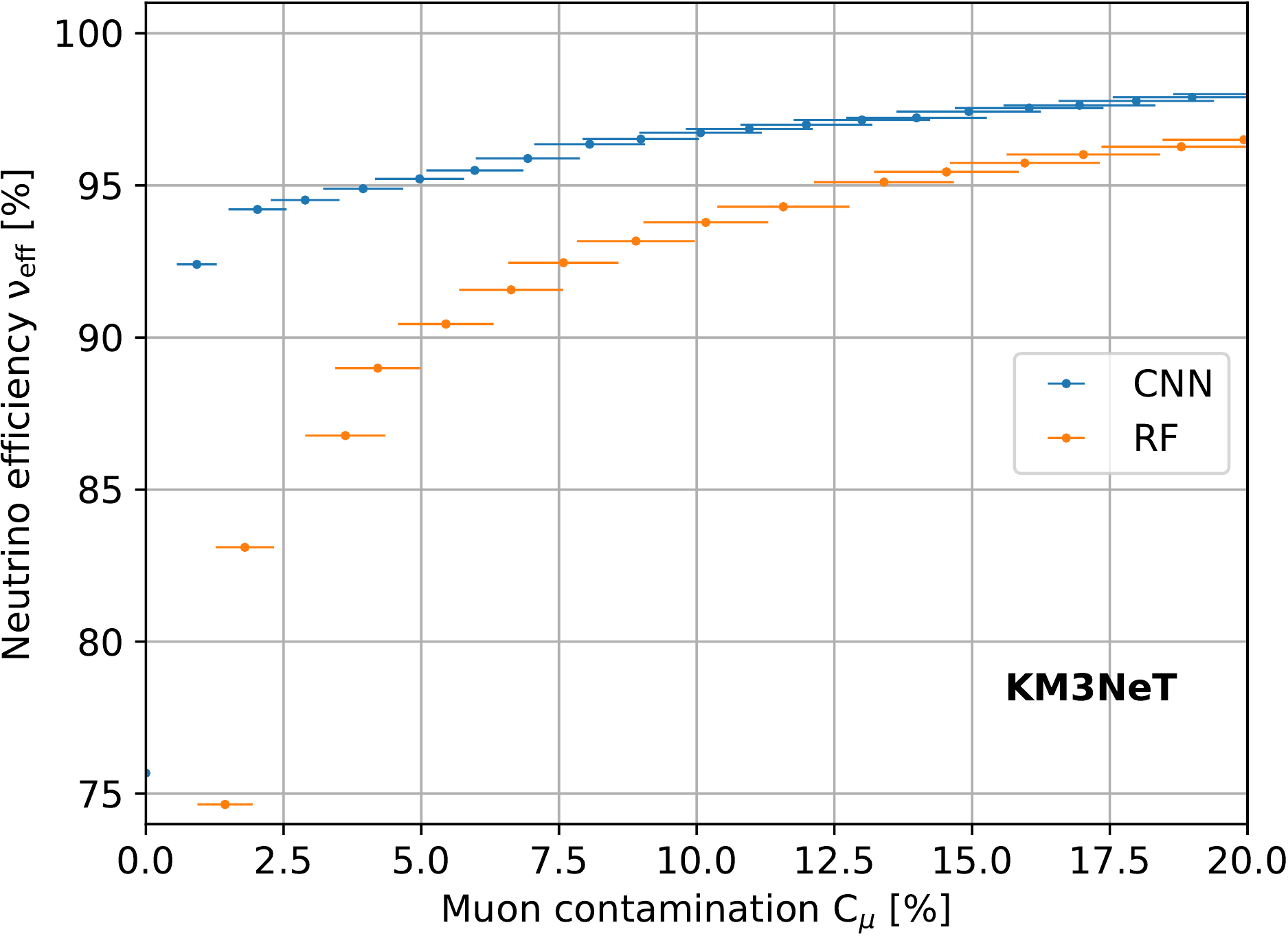}
	\caption{Neutrino efficiency, $\upnu_{\text{eff}}$, versus
          atmospheric muon event contamination, $\text{C}_{\upmu}$,
          weighted with the Honda atmospheric neutrino flux
          \autocite{HKKM} model. The CNN (RF) performance is depicted
          in blue (orange). Only neutrinos with a MC energy in the
          range of \SIrange{10}{20}{\giga\electronvolt} have been
          used.}
	\label{fig:muon_cont_vs_neutr_eff_10_20}
\end{figure}

Comparing different neutrino energy ranges,
\SIrange{1}{5}{\giga\electronvolt} in
\myfref{fig:muon_cont_vs_neutr_eff_1_5} and
\SIrange{10}{20}{\giga\electronvolt} in
\myfref{fig:muon_cont_vs_neutr_eff_10_20}, it can be seen that the
performance gap between the CNN and the RF classifier widens with
increasing neutrino energy. 
A possible explanation may be that small details in the distribution
of the measured hits increase in importance if the neutrino
events are less energetic and thus produce less hits. Then, the
limitations of the input images, which do not contain the full
information about an event, may become more relevant compared to
events with higher energies and many signal hits.

For random noise events, the performance of the CNN and the RF
classifier are comparable. In particular, both methods achieve about
99\% neutrino efficiency at 1\% random noise event contamination. As
expected, the suppression of atmospheric muon events is significantly
more difficult.
 


\section{Event topology classifier}
\label{sect:event_topology_classifier}
Similar to the background classifier, a RF is used in the
current
KM3NeT/ORCA analysis pipeline to separate track-like from shower-like
neutrino events.  This section introduces a CNN-based event topology
classifier that distinguishes between these two event types.

\subsection{Image generation}

The input event images for the CNN-based track--shower classifier are
similar to the ones of the background classifier introduced in
\mysref{subsect:bg_image_gen}, i.e. the input also consists of XYZ-T/P images. The timecut for the hit selection is tighter
with respect to the background classifier. Since background events
have already been rejected by the background classifier, the data
presented to the track--shower classifier are mostly neutrino
interactions which produce secondary particles that, on average,
traverse smaller parts of the detector compared to atmospheric muon
events. The event class that shows the broadest signal hit time
distribution is the class of $\numuCC$ events due to the outgoing
muon. Therefore, the timecut of the track--shower classifier is set
based on these events. The time distribution of signal hits relative
to the mean time of the triggered hits for $\numuCC$ events is shown
in \myfref{fig:timecut_muon-CC}. Based on this distribution the
timecut is set to an interval of [$-\SI{250}{\nano\second},
  +\SI{500}{\nano\second}$], as indicated by the dashed, black lines
in \myfref{fig:timecut_muon-CC}.

Since the timecut interval is smaller than the one used for the
background classifier, less time bins (60) are used for the time
dimension.
This implies a reduction of the time resolution of about 30\% with
respect to the background classifier, i.e. \SI{12.5}{\nano\second} per
time bin.  The light emission profile of hadronic and electromagnetic
showers in the GeV range has an extension of at most a few metres, see
Fig.~70 in Ref.~\autocite{Adrian-Martinez2016}.  Since a muon of comparable
energy induces the emission of Cherenkov radiation along a
significantly greater path length, the reduced time binning still
provides a sufficient resolution to distinguish a shower-like from a
track-like event topology, while significantly speeding up the
training of the CNN.  Consequently, the XYZT images now have
$\text{11} \times \text{13} \times \text{18} \times \text{60}$ pixels.

\subsection{Network architecture}
\label{subsect:ts_network_arch}

The network architecture, as depicted in
\mytref{tab:ts_model_xyztc_1}, is the same as in
\mysref{subsect:bg_network_arch}, except for additional dropout
layers. Since the size of the training dataset is significantly
smaller than that for the background classifier, overfitting can be
observed without any regularisation. Thus, dropout layers with a rate
of $\delta = 0.1$ are added in every convolutional block and also in between the
last two fully connected layers.

\renewcommand{\arraystretch}{1.1}
\begin{table}[h!]
	\centering
	\setlength{\extrarowheight}{0.1cm}
	\setlength{\tabcolsep}{0.4cm}
	\begin{tabular}{|c|rcl|}
		
		\multicolumn{1}{c}{Building block / layer} & \multicolumn{3}{c}{Output dimension} \\ 
		\hline
		\rowcolor{Gray} XYZ-T Input & 11 $\times$ 13 $\times$ 18 & $\times$ & 60 \\ 
		\rowcolor{Gray} XYZ-P Input & 11 $\times$ 13 $\times$ 18 & $\times$ & 31 \\ 
		\rowcolor{Gray} Final stacked XYZ-T + XYZ-P Input & 11 $\times$ 13 $\times$ 18 & $\times$ & 91 \\ \hline
		Convolutional block 1 (64 filters, $\delta=\text{0.1}$)  & 11 $\times$ 13 $\times$ 18 & $\times$ & 64 \\ \hdashline
		Convolutional block 2 (64 filters,  $\delta=\text{0.1}$) & 11 $\times$ 13 $\times$ 18 & $\times$ & 64 \\ \hdashline
		Convolutional block 3 (64 filters,  $\delta=\text{0.1}$) & 11 $\times$ 13 $\times$ 18 & $\times$ & 64 \\ \hdashline
		Convolutional block 4 (64 filters,  $\delta=\text{0.1}$) & 11 $\times$ 13 $\times$ 18 & $\times$ & 64 \\ \hdashline
		Convolutional block 5 (64 filters,  $\delta=\text{0.1}$) & 11 $\times$ 13 $\times$ 18 & $\times$ & 64 \\ \hdashline
		Convolutional block 6 (64 filters,  $\delta=\text{0.1}$) & 11 $\times$ 13 $\times$ 18 & $\times$ & 64 \\ \hdashline
		Max pooling (2,2,2) & 5 $\times$ 6 $\times$ 9 & $\times$ & 64 \\ \hline
		
		Convolutional block 1 (128 filters,  $\delta=\text{0.1}$) & 5 $\times$ 6 $\times$ 9 & $\times$ & 128 \\ \hdashline 
		Convolutional block 2 (128 filters,  $\delta=\text{0.1}$) & 5 $\times$ 6 $\times$ 9 & $\times$ & 128 \\ \hdashline 
		Convolutional block 3 (128 filters,  $\delta=\text{0.1}$) & 5 $\times$ 6 $\times$ 9 & $\times$ & 128 \\ \hdashline 
		Convolutional block 4 (128 filters,  $\delta=\text{0.1}$) & 5 $\times$ 6 $\times$ 9 & $\times$ & 128 \\ \hdashline 
		Max pooling (2,2,2) &  2 $\times$ 3 $\times$ 4 & $\times$ & 128 \\ \hline
		
		Flatten & \multicolumn{3}{c|}{3072}  \\ \hdashline 
		Dense + ReLU & \multicolumn{3}{c|}{128} \\ \hdashline 
		Dropout ( $\delta=\text{0.1}$ ) & \multicolumn{3}{c|}{128} \\ \hdashline 
		Dense + ReLU \quad & \multicolumn{3}{c|}{32} \\ \hdashline 
		Dense + Softmax & \multicolumn{3}{c|}{2} \\ \hline
		
	\end{tabular}
	\vspace{2mm}
	\caption{Network structure of the track--shower classifier
          three-dimensional CNN model with XYZ-T/P input. The symbol
          $\delta$ specifies the dropout rate used in the respective
          convolutional block, cf.~\mysref{sec:cnns}.}
	\label{tab:ts_model_xyztc_1}
\end{table}
\renewcommand{\arraystretch}{1.0}

\subsection{Preparation of training, validation and test data}
\label{subsec:ts_classifier:preparation_data}

In order to train the CNN track--shower classifier, only simulated
neutrino events are used. The total neutrino dataset is rebalanced such
that 50\% of the events are track-like ($\numuCC$) and 50\% are
shower-like. The shower class consists of 50\% $\nueCC$ and 50\%
$\nueNC$. Additionally, the dataset has been balanced in such a way
that the ratio of track-like to shower-like events is always one, independent of neutrino energy.

The rebalanced dataset is then split into three datasets with 70\%
training, 6\% validation, and 24\% test events.  In total, the
training dataset contains about $\text{1.3}\times\text{10}^{\text{7}}$
events.

Using a Nvidia Tesla V100 GPU, it takes about one and a half weeks to
fully train this CNN track--shower classifier.

\subsection{Performance and comparison to Random Forest classifier}

\begin{figure}[h!]
  \centering
\begin{overpic}[width=0.675\linewidth,page=1]{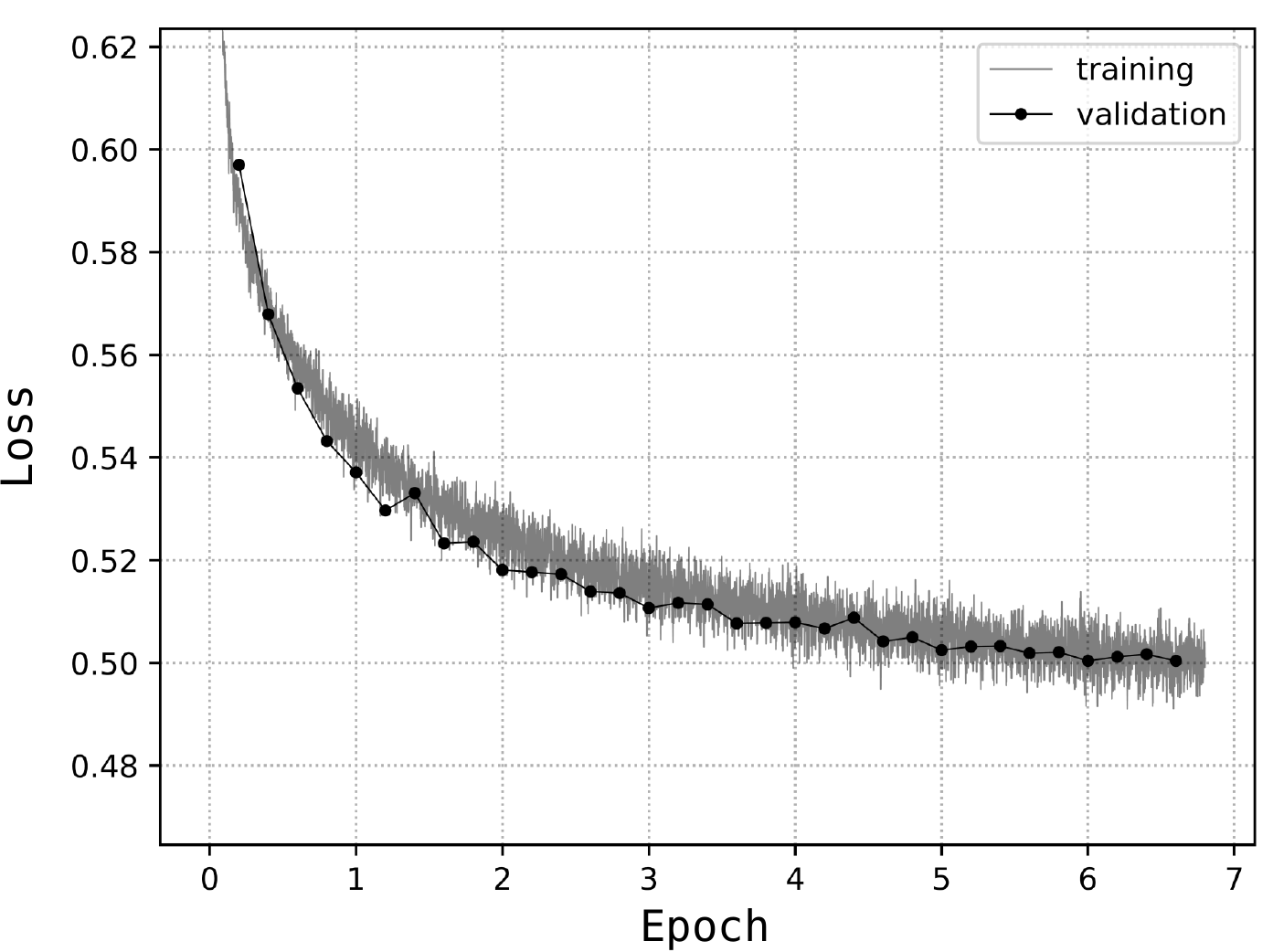}
\put (40,66) {\textsf{KM3NeT}}
\end{overpic}
	\caption{Training (grey lines) and validation (black circles)
          cross-entropy loss of the track--shower classifier during the
          training process.  Each line element of the training loss
          represents an average over 250 batches,
          i.e. $\text{250}\times\text{64}$ event images.}
	\label{fig:ts_loss}
\end{figure}

The evolution of the cross-entropy loss \autocite{Goodfellow2016}
during the training is shown in \myfref{fig:ts_loss}.
Even though the validation cross-entropy loss is within the
fluctuations of the training loss curve at the end of the training,
some minor overfitting occurs.  The reason is that during the
application of the trained network on the validation dataset, no
neurons are dropped by the dropout layers, contrary to the training
phase.
Therefore, the validation loss should be lower than the
average training loss, if no overfitting is observed.
This can 
be seen by investigating the training and validation loss curves in
the earlier stages of the training process, e.g. between epoch 0 and
3. Here, the validation loss is typically found at the bottom of the
training loss curve.

\begin{figure}[ht!]
	\centering
        \includegraphics[width=0.675\textwidth]{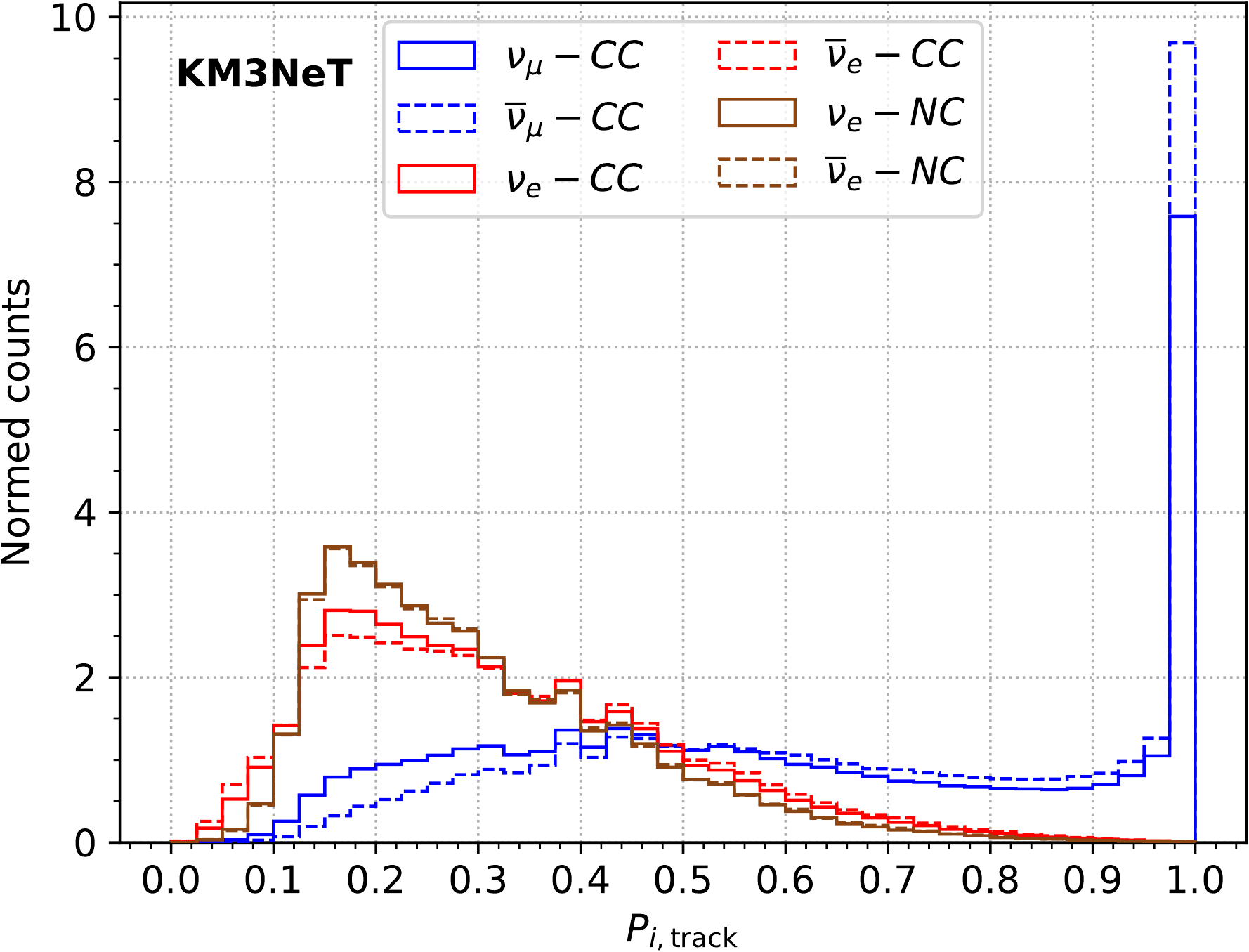}
	\caption{
          Distribution of the CNN track probability
          $P_{i,\text{track}}$ for pre-selected neutrino events of
          different flavours and interaction channels in the energy
          range of \SIrange{1}{40}{\giga\electronvolt}. All
          distributions have been normalised to the area under each
          histogram.}
	\label{fig:ts_prob_track}
\end{figure}

\begin{figure}[h!]
	\centering
        \includegraphics[width=0.675\textwidth]{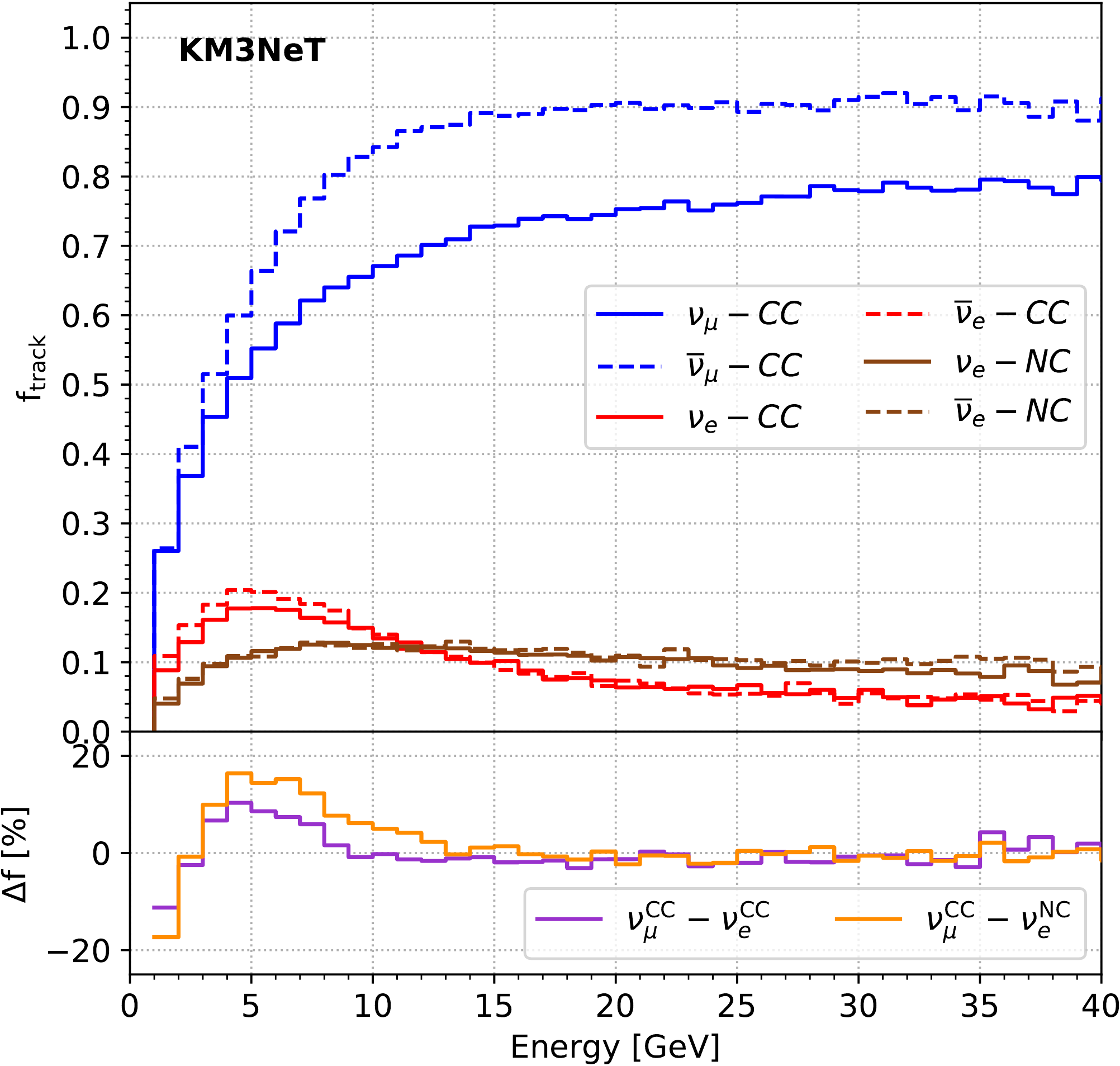}
	\caption{Fraction of events classified as track-like,
          $\text{f}_{\text{track}}$, with a CNN track probability
          $P_{i,\text{track}}>\text{0.5}$, for different interaction
          channels (top panel) versus the true MC neutrino energy. The
          relative improvement $\Delta\text{f}$,
          cf.~\myeref{eq:frel}, of the CNN with respect to the RF
          classifier in discriminating between $\numuCC$ events and
          shower-like $\nueCC$/ $\nueNC$ channels is shown in the
          bottom panel. Here, the contributions of neutrinos and
          antineutrinos are averaged for each flavour and
          interaction. The discrimination power is defined as the
          difference between the fractions of events classified as
          track-like and shower-like in the top panel.
        }
	    \label{fig:ts_perf_track}
\end{figure}

The binned probability distribution for all used neutrino events with
energies in the range of \SIrange{1}{40}{\giga\electronvolt} to be
classified as track-like is shown in \myfref{fig:ts_prob_track}. This
energy range has been chosen since the classification performance
saturates at about \SI{40}{\giga\electronvolt}, as can be seen in
\myfref{fig:ts_perf_track} and further discussed below. The classified
neutrino events have been selected according to the criteria described
in \mysref{subsect:bg_performance}.
\indent About 25\% of $\numuCC$ and $\anumuCC$ events are identified as
track-like events with a probability close to one. The correct
identification of muon tracks increases with their length and hence
with their energy. As the outgoing muon has on average higher energy
for $\anumuCC$ than for $\numuCC$ events, $\anumuCC$ events have a
higher probability to be identified as track-like.

The top panel of \myfref{fig:ts_perf_track} shows the fraction of
events classified as track-like as a function of the neutrino energy
in the range of \SIrange{1}{40}{\giga\electronvolt}. An event is
accepted if its CNN probability to be a track-like event is 0.5 or above.
The fraction of correctly classified $\numuCC$ and $\anumuCC$ events
rises strongly with neutrino energy. The reason is that the spatial
and temporal distribution of hits induced by low-energy $\numuCC$
events is similar to that of shower-like events, since the outgoing
muon stops at these energies after propagating only a few meters in
the detector.  Comparing $\nueCC$ and $\anueCC$ with energies below
\SI{10}{\giga\electronvolt}, it can be seen that $\anueCC$ are more
likely classified as track-like by the CNN classifier. This is again
due to the fact that the outgoing charged lepton carries on average
more energy in $\anuCC$
interactions with respect to $\nuCC$
interactions. On the other hand, for energies below
\SI{10}{\giga\electronvolt}, NC interactions have a lower
misclassification rate compared to $\nueCC$ interactions, since there
is no charged secondary lepton that could mimic the signature of a
low-energy muon as induced in $\numuCC$ events. For energies above
\SI{10}{\giga\electronvolt} the opposite is true, since a localised
and bright light source disfavours a $\numuCC$ event.

A measure for the separation power of the classifier is the difference
between the fractions of recognised track-like events for $\numuCC$
($\text{f}^{\numuCC}_{\text{track}}$, blue line) and $\nue$
($\text{f}^{\nueCCNC}_{\text{track}}$, red and brown lines) in the top
panel of \myfref{fig:ts_perf_track}. In the bottom panel the ratio of
these differences, as determined by the CNN and RF classifiers,
i.e. the relative improvement $\Delta\text{f}$ of the CNN with respect
to the RF classifier, is shown for different channel combinations:
\begin{equation}
\Delta\text{f} = \left(\text{f}^{\numuCC}_{\text{track}}-\text{f}^{\nueCCNC}_{\text{track}}\right)_\text{CNN} \;/\; \left(\text{f}_{\text{track}}^{\numuCC} -\text{f}_{\text{track}}^{\nueCCNC}\right)_\text{RF}\,.
\label{eq:frel}
\end{equation}
%
It can be seen that the separation power between track-like and
shower-like events depends on the neutrino energy.  The CNN performs
better than the RF in the energy range of roughly
\SIrange{3}{10}{\giga\electronvolt}, while the RF performs better than
the CNN for energies below \SI{2}{\giga\electronvolt}.  As explained
at the end of \mysref{subsect:bg_performance}, this could again be due
to limitations in the input images that become more relevant for
events with only a few hits. Above \SI{15}{\giga\electronvolt} the
separation power of the two classifiers is roughly equal.

Since the performance comparison shown in \myfref{fig:ts_perf_track}
depends on the threshold value that is set for an event to be
classified as track-like, another comparison metric
independent of the threshold value is introduced.
$P_{\text{track}}^{\numu}$ is defined as the probability 
that a classifier recognises a $\numuCC$ event as track-like, and 
similarly for $\nueCC$.
Based on \myfref{fig:ts_prob_track}, the
ability of a classifier to separate track-like and shower-like events
can be estimated by calculating
the energy-dependent correlation factor $c(E)$ of the two binned probability distributions
$P_{i, \text{track}}^{\numu}(E)$ and $P_{i, \text{track}}^{\nue}(E)$.
The \textit{separability} $S(E)$ is then given by:

\begin{equation}
S(E) = 1 - c(E) = 1 - \frac{\sum_{i} P^{\numu}_{i,
    \text{track}}(E) \cdot P^{\nue}_{i, \text{track}}(E)}
    { \sqrt{\sum_{i} \left( P^{\numu}_{i, \text{track}}(E)
    \right) ^ 2 \cdot \sum_{i} \left( P^{\nue}_{i,
      \text{track}}(E) \right) ^ 2 } }\,,
\end{equation}
%
where the index $i$ in $P^{\numu/\nue}_{i, \text{track}}$ is the $i$-th bin
of the distribution $P_{\text{track}}^{\numu/\nue}$ as shown in
\myfref{fig:ts_prob_track}.

\begin{figure}[h!]
	\centering
        \includegraphics[width=0.675\textwidth]{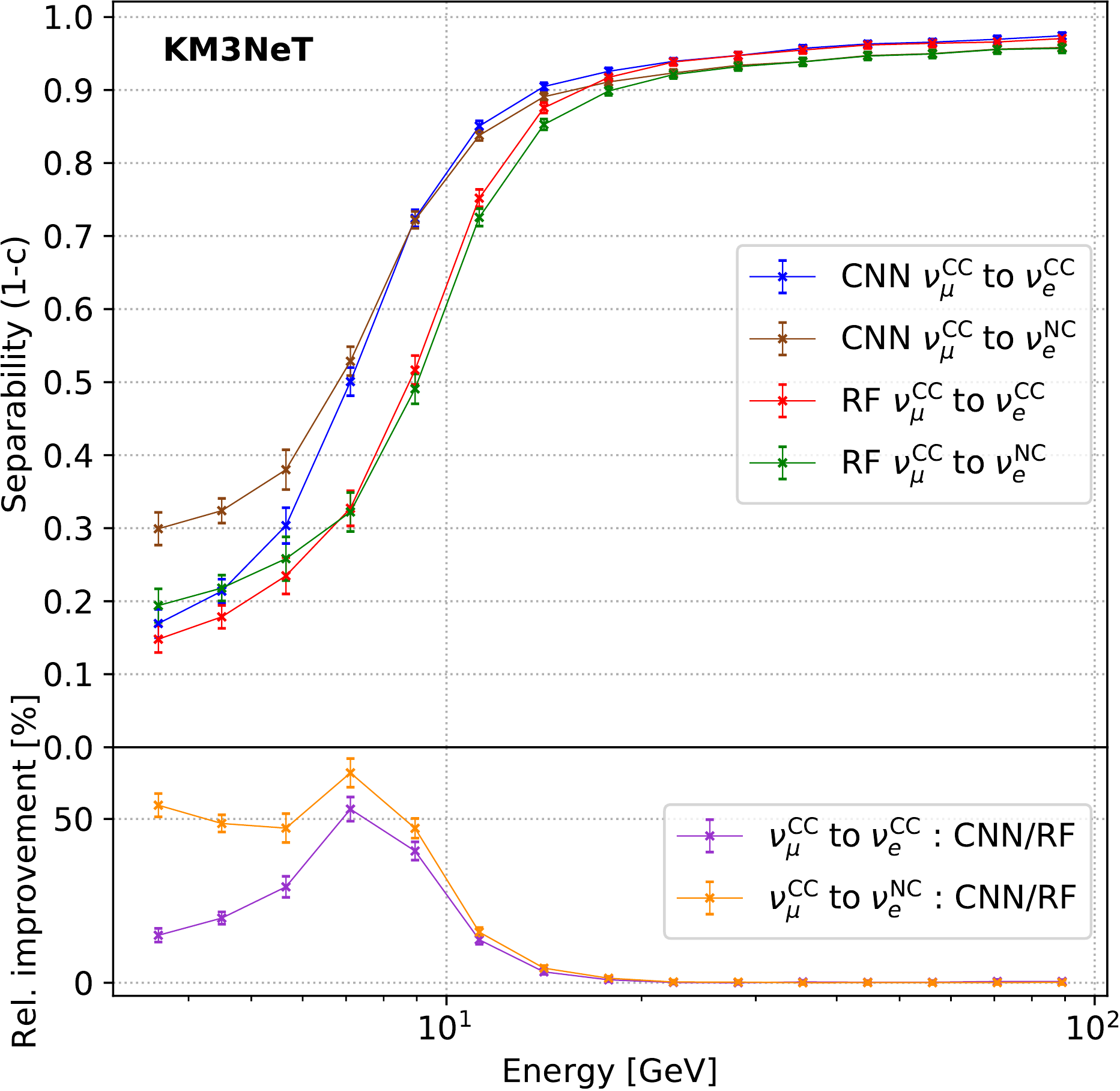}
	\caption{Separability $S(E) = 1 - c(E)$ based on $\numuCC$ and
          $\nueCC$ ($\numuCC$ and $\nueNC$) events as a function of
          true MC neutrino energy for the CNN represented by blue
          (brown) markers and the RF represented by red (green)
          markers (top panel). Only $\upnu$ (not $\anu$) have been
          used for the comparison.  The relative improvement of the
          CNN with respect to the RF classifier in discriminating
          between $\numuCC$ events and shower-like $\nueCCNC$
          channels is shown in the bottom panel. The error bars
          indicate statistical errors only.}
	\label{fig:ts_separability}
\end{figure}

The separability $S(E)$ as a function of binned neutrino energy for
the CNN and RF classifier is shown in \myfref{fig:ts_separability}.
It is evident that the CNN-based classifier performs better by an
absolute value of up to about 20\% for energies below
\SI{10}{\giga\electronvolt}.  For these energies, the RF- and
CNN-based classifier show both a better separability of $\numuCC$ and
$\nueNC$ compared to $\nueCC$ events, although this loss in
separability is less pronounced and reverses above
\SI{7}{\giga\electronvolt} for the RF classifier, i.e. at slightly
lower energies than for the CNN classifier.


\section{Event regression}
\label{sect:event_regressor}

This section introduces the CNN-based \textit{event regressor}. The
designed CNN now reconstructs continuous properties of the interacting
neutrinos such as the direction, energy, the position of the
interaction, i.e. the vertex, and an estimate for the uncertainty of
each of these inferred observables. In the reconstruction pipeline for
KM3NeT/ORCA as described in Ref.~\autocite{Adrian-Martinez2016}, these
reconstruction tasks are handled by two maximum-likelihood-based
algorithms optimised for track-like and shower-like events,
respectively.

The major difference of the CNN for the regression task with respect
to the classifiers in the previous sections is that its output is a
continuous instead of a categorical variable.

\subsection{Image generation}

For the event regressor, the event images are created in a similar way
to the image generation for the background and the track--shower
classifier. Thus, both XYZT and XYZP images are produced and then
stacked in the last dimension.  Since the background classifier has
already been applied to the data, at this final stage of the
reconstruction pipeline neutrinos remain predominantly as input to the
event regressor. Hence, exactly the same images as specified in
\mysref{sect:event_topology_classifier} are used as input for the CNN
event regressor.

\subsection{Network architecture and loss functions}
\label{sec:regression_arch_loss}
The network architecture of the CNN for the event regressor is similar
to the one used for the background and track--shower
classifier. However, for this work, it has been found that dropout
layers in the CNN event regressor increase fluctuations in the
training loss, even with a small dropout rate of
$\delta=\text{0.05}$. Furthermore, the validation loss of the event
regressor CNNs with dropout layers has proven to be significantly
worse than for CNNs without dropout. Thus, no dropout layers are
included.  The fully-connected layers after the convolutional layers
are also different. The properties that should be reconstructed by the
network are the energy, the direction, and the vertex position of the
initial neutrino interaction. For the direction and the vertex, the
CNN output consists of two arrays with three elements each, containing
the components of the vertex position vector and the direction
cosines, respectively. Consequently, the output of the network
consists of seven reconstructed floating-point numbers, denoted as
$\yrecovec$, one component for the energy, three for the direction and
three for the vertex position.
The final fully-connected output layer therefore consists of seven
neurons.  The CNN architecture up to this point is shown in
\mytref{tab:regr_model_xyztc_1}.
\renewcommand{\arraystretch}{1.1}
\begin{table}[h!]
	\centering
	\setlength{\extrarowheight}{0.1cm}
	\setlength{\tabcolsep}{0.4cm}
	\begin{tabular}{|c|rcl|}
		
		\multicolumn{1}{c}{Building block / layer} & \multicolumn{3}{c}{Output dimension} \\ 
		\hline
		\rowcolor{Gray} XYZ-T Input & 11 $\times$ 13 $\times$ 18 & $\times$ & 60 \\ 
		\rowcolor{Gray} XYZ-P Input & 11 $\times$ 13 $\times$ 18 & $\times$ & 31 \\ 
		\rowcolor{Gray} Final stacked XYZ-T + XYZ-P Input & 11 $\times$ 13 $\times$ 18 & $\times$ & 91 \\ \hline
		
		Convolutional block 1 (64 filters) & 11 $\times$ 13 $\times$ 18 & $\times$ & 64 \\ \hdashline
		Convolutional block 2 (64 filters) & 11 $\times$ 13 $\times$ 18 & $\times$ & 64 \\ \hdashline
		Convolutional block 3 (64 filters) & 11 $\times$ 13 $\times$ 18 & $\times$ & 64 \\ \hdashline
		Convolutional block 4 (64 filters) & 11 $\times$ 13 $\times$ 18 & $\times$ & 64 \\ \hdashline
		Convolutional block 5 (64 filters) & 11 $\times$ 13 $\times$ 18 & $\times$ & 64 \\ \hdashline
		Convolutional block 6 (64 filters) & 11 $\times$ 13 $\times$ 18 & $\times$ & 64 \\ \hdashline
		Max pooling (2,2,2) & 5 $\times$ 6 $\times$ 9 & $\times$ & 64 \\ \hline
		
		Convolutional block 1 (128 filters) & 5 $\times$ 6 $\times$ 9 & $\times$ & 128 \\ \hdashline 
		Convolutional block 2 (128 filters) & 5 $\times$ 6 $\times$ 9 & $\times$ & 128 \\ \hdashline 
		Convolutional block 3 (128 filters) & 5 $\times$ 6 $\times$ 9 & $\times$ & 128 \\ \hdashline 
		Convolutional block 4 (128 filters) & 5 $\times$ 6 $\times$ 9 & $\times$ & 128 \\ \hdashline 
		Max pooling (2,2,2) &  2 $\times$ 3 $\times$ 4 & $\times$ & 128 \\ \hline
		
		Flatten & \multicolumn{3}{c|}{3072}  \\ \hdashline 
		Dense + ReLU & \multicolumn{3}{c|}{128} \\ \hdashline 
		Dense + ReLU \quad & \multicolumn{3}{c|}{32} \\ \hdashline 
		Dense + Linear & \multicolumn{3}{c|}{7} \\ \hline
		
	\end{tabular}
	\vspace{2mm}
	\caption{Network structure of the CNN event regressor with
          XYZ-T/P input, but without the additional fully-connected
          network for the uncertainty estimation. In the convolutional
          blocks and the last fully-connected layers no dropout layer is
          used.}
	\label{tab:regr_model_xyztc_1}
\end{table}
\renewcommand{\arraystretch}{1.0}

Since the aim of this network is to reconstruct continuous variables,
a categorical cross-entropy loss \autocite{Goodfellow2016}, as for the
background and track--shower classifier, cannot be used. Instead, other
cross-entropy-based loss functions can be employed, e.g. the mean squared
error (MSE) or the mean absolute error (MAE)
\autocite{Goodfellow2016}, which are defined as:
\begin{equation}
\text{MSE} = \frac{1}{n} \sum_{i=1}^{n} \left( \yreco - \ytrue \right)^2,
\end{equation}
\begin{equation}
\text{MAE} = \frac{1}{n} \sum_{i=1}^{n} \left| \yreco - \ytrue \right|.
\end{equation}
The main difference between the MSE and the MAE loss is that the MSE
loss penalises outlier events that are badly reconstructed with a
significantly larger loss value compared to the MAE loss. Since many
detectable events are not fully contained in the KM3NeT/ORCA detector,
these events are on average reconstructed more poorly. Thus, the MAE
loss is used for the energy, direction and vertex estimation, so that
the network focuses less on outlier events.\\
\indent Using a neural network,
it is also possible to estimate the uncertainties on the seven
components of $\yrecovec$.  One possibility is to let the network
predict the absolute residual of the MAE loss. Thus, one needs an
additional neuron that yields the estimated uncertainty,
$\upsigma_\text{reco}$, for each of the reconstructed components of
$\yrecovec$.  Consequently, the loss function that measures the
quality of the estimated uncertainty value is:
\begin{equation}
L = \frac{1}{n}\sum_{i=1}^{n} \left( \upsigma_\text{reco} - \left|
\ytrue - \yreco \right| \right)^2.
\end{equation}
Using this loss function $L$, the network learns to estimate the
average absolute residual:
\begin{equation}
\upsigma_\text{reco} \approx \left\langle \left|
\ytrue - \yreco \right| \right\rangle .
\end{equation}
Assuming that the residuals are normally distributed, the estimated
uncertainty value $\upsigma_\text{reco}$ can be converted to a standard
deviation by multiplying by
$\sqrt{\frac{\pi}{2}}$ \autocite{10.1093/biomet/27.3-4.310}.

\begin{figure}[h!]
	\centering
        \includegraphics[width=0.675\textwidth]{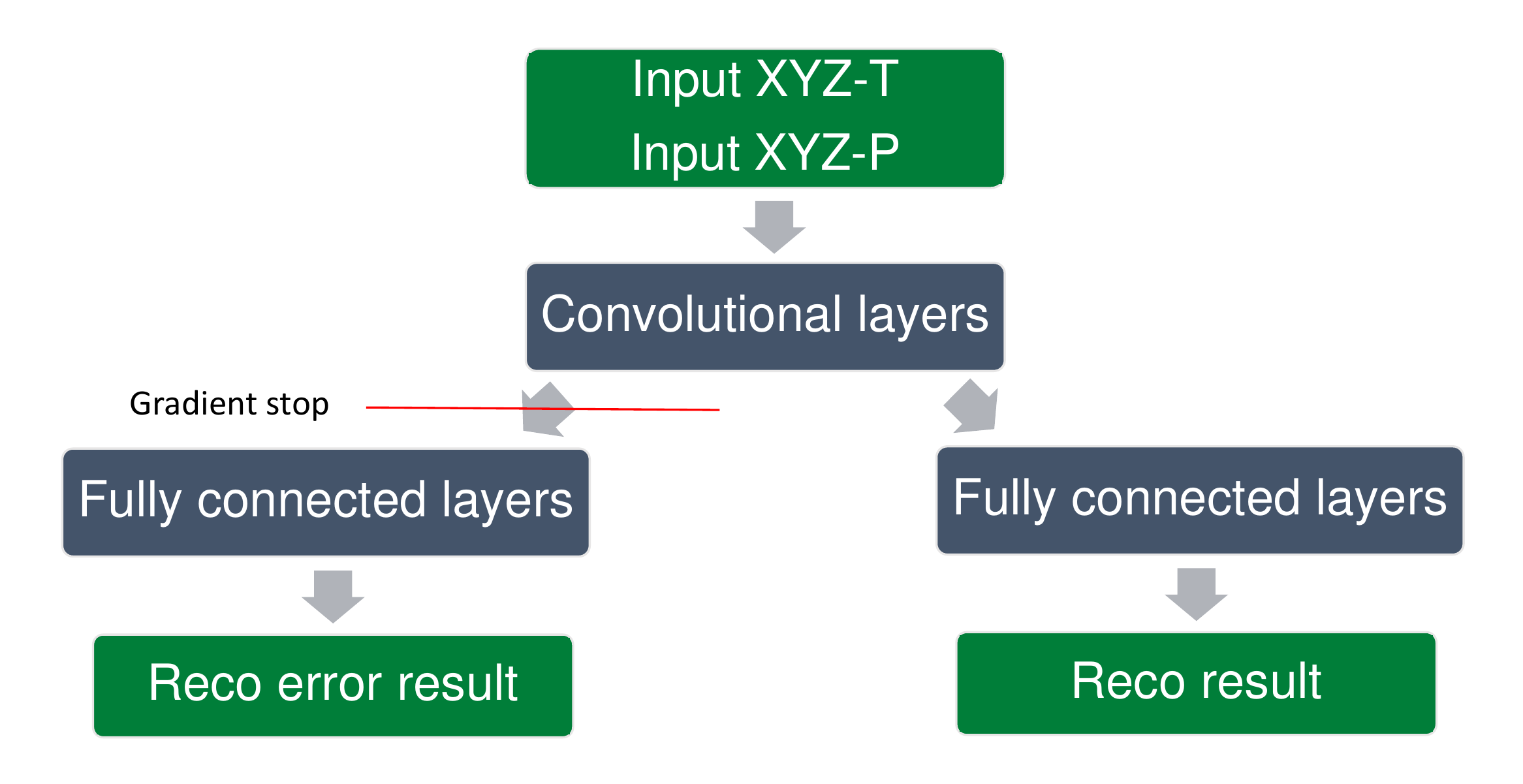}
	\caption{Scheme of the CNN architecture used for the event
          regressor. The properties of the convolutional layers and of
          the fully-connected network for the reconstruction of
          $\yrecovec$ is depicted in \mytref{tab:regr_model_xyztc_1} and
          in \mytref{tab:regr_model_xyztc_errors}.}
	\label{fig:error_net_scheme}
\end{figure}

In general, it can be expected that features that are learned by the
convolutional part of the network and that are useful for the
reconstruction are also crucial for the inference of the uncertainty
on the reconstruction.  Therefore, both inference tasks are handled by
using the same CNN.  The CNN, however, must be prevented from focusing
too much on the uncertainty estimation in the learning process, since
its main goal is the prediction of $\yrecovec$ and not of the
uncertainties. For this purpose, a second fully-connected network with
three layers is added after the convolutional output, but the gradient
update during the back-propagation stage is stopped after the first
fully-connected layer of this sub-network,
as shown in \myfref{fig:error_net_scheme}.
The specific structure of the fully-connected network for the uncertainty 
reconstruction estimation is shown in \mytref{tab:regr_model_xyztc_errors}.
\renewcommand{\arraystretch}{1.1}
\begin{table}[h!]
	\centering
	\setlength{\extrarowheight}{0.1cm}
	\setlength{\tabcolsep}{0.4cm}
	\begin{tabular}{|c|rcl|}
		
		\multicolumn{1}{c}{Building block / layer} & \multicolumn{3}{c}{Output dimension} \\ 
		\hline
		
		Dense + ReLU & \multicolumn{3}{c|}{128} \\ \hdashline 
		Dense + ReLU \quad & \multicolumn{3}{c|}{64} \\ \hdashline
		Dense + ReLU \quad & \multicolumn{3}{c|}{32} \\ \hdashline 
		Dense + Linear & \multicolumn{3}{c|}{7} \\ \hline
		
	\end{tabular}
	\vspace{2mm}
	\caption{Layout of the fully-connected uncertainty
          reconstruction network, which is part of the event
          regressor CNN with XYZ-T/P input. The input to this dense
          network is the flattened output of the last convolutional
          layer of the main CNN.}
	\label{tab:regr_model_xyztc_errors}
\end{table}
\renewcommand{\arraystretch}{1.0}

\subsection{Preparation of training, validation and test data}
In order to train the CNN event regressor, again only simulated
neutrino events are used. Similar to the track--shower classifier, the
dataset consists of 50\% track-like events ($\numuCC$) and 50\%
shower-like events (25\% $\nueCC$, 25\% $\nueNC$). Additionally, the
total dataset has been split again into a training, validation and
test dataset. As for the track--shower classifier, the training data
makes up 70\%, the validation dataset 6\% and the test dataset 24\% of
the total rebalanced dataset.  The track and shower classes have not
been balanced to be independent of neutrino energy,
cf.~\mysref{subsec:ts_classifier:preparation_data}, in order to get a
larger training dataset compared to the track--shower classifier. In
total, the training dataset then contains about
$\text{14}\times\text{10}^{\text{6}}$ events.

The MC energy of $\nueNC$ events is set to the visible
energy during the training, since the network cannot infer how much
energy is carried away by the outgoing neutrino in the neutral current
interaction.

Using a Nvidia Tesla V100 GPU, it takes about one and a half weeks to
fully train this CNN event regressor.

\subsection{Loss functions and loss weights}

Before training the CNN event regressor, suitable \textit{loss
  weights} need to be defined in order to scale the losses to the same
magnitude. For example, the neutrino energy in the dataset ranges from
\SIrange{1}{100}{\giga\electronvolt}, while the components of the
direction vector range from \SIrange{-1}{1}{}. Thus, the impact of the
energy loss would be overwhelmingly large compared to the loss that is
contributed by the direction reconstruction.  Consequently, one needs
to scale the individual losses of the components of $\yrecovec$ and their
estimated uncertainties with a loss weight, such that all single
losses are of the same order of magnitude at the start of the
training. However, it has been observed that the CNN does not start to
learn during the first few epochs of the training, if the individual
losses of the energy, direction and vertex are of the same scale. The
reason is that direction and vertex contribute six out of seven
variables to be learned so that their impact on the total loss is
dominant. Hence, any improvement with respect to the energy
reconstruction, i.e. in the discrimination of background and signal
hits, is marginalised by random statistical fluctuations in the loss
for the direction or vertex components.

It is expected that the first
concept the neural network will learn during the training is to
distinguish signal and background hits. At the same time, the
recognition of signal hits is a prerequisite for the direction and
vertex reconstruction.  Thus, the loss weights have been set in such a
way that the scale of the energy loss at the start of the training is
about two times larger than the scale of the three combined
directional losses. For the same reason, the loss weights of the
vertex reconstruction are set in such a way that they have the same
scale as the directional losses.  Indeed, it could be observed that
the energy loss always converges significantly earlier during the
training than the loss for the direction or vertex.  Since the scale
of the individual losses changes dramatically after the network has
started to learn a property such as the energy or the direction, the
loss weights need to be tuned during the training.

As a general strategy, it has proven useful to do this after the
energy loss has sufficiently converged. How often and by how much the
individual loss weights are retuned is a parameter optimisation
process.
For this
work, the loss weights of the direction and vertex components have
been increased by a factor of three after the energy loss convergence,
and, at the end of the training, these loss weights have been doubled
again to investigate if further improvements would occur, which was
not the case.
Increasing the weights for the direction and vertex losses did not
significantly worsen the energy loss.

For the loss weights of the uncertainty outputs none of this matters,
since their gradients are stopped after the first fully-connected
layer of the uncertainty reconstruction sub-network. Thus, the loss
weights for these variables just need to be set such that the scale of
the individual losses is about the same.

\subsection{Energy reconstruction performance}
\label{sec:regression_energy}

In this section the neutrino energy reconstruction performance of the
CNN event regressor is investigated.  The study and the comparisons to
the maximum-likelihood-based reconstruction algorithm, described in
Ref.~\autocite{Adrian-Martinez2016} and called \textit{SOR} in the
following, focus on shower-like events.

Events have been selected according to the criteria detailed in
\mysref{subsect:bg_performance}. In particular, only events
reconstructed as up-going by the SOR algorithm are selected for the
comparison and, additionally, the vertex position, direction, and
energy of true $\nueCC$ events must have been reconstructed with high
confidence by SOR.

\begin{figure}[h!]
	\centering
        \includegraphics[width=0.675\textwidth,page=1]{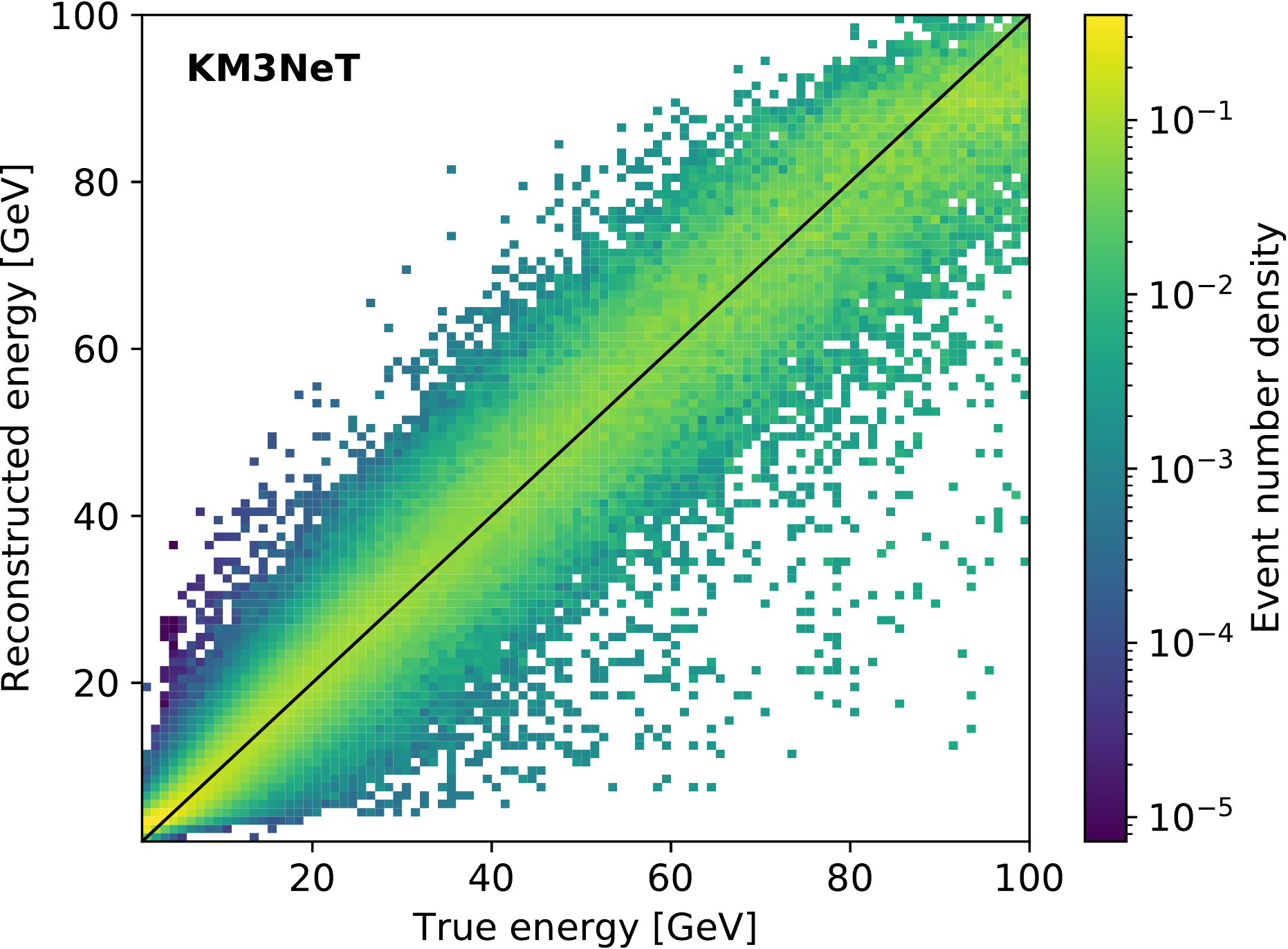}
	\caption{Energy reconstructed by the CNN regressor versus true MC
          neutrino energy for reconstructed $\nueCC$ events. The
          distribution has been normalised to unity in each true
          energy bin.}
	\label{fig:regr_2d_energy_nueCC_not_corrected}
\end{figure}

\begin{figure}[h!]
\centering
\begin{minipage}[c]{0.48\textwidth}
\centering
    {\includegraphics[width=\textwidth, page=1]{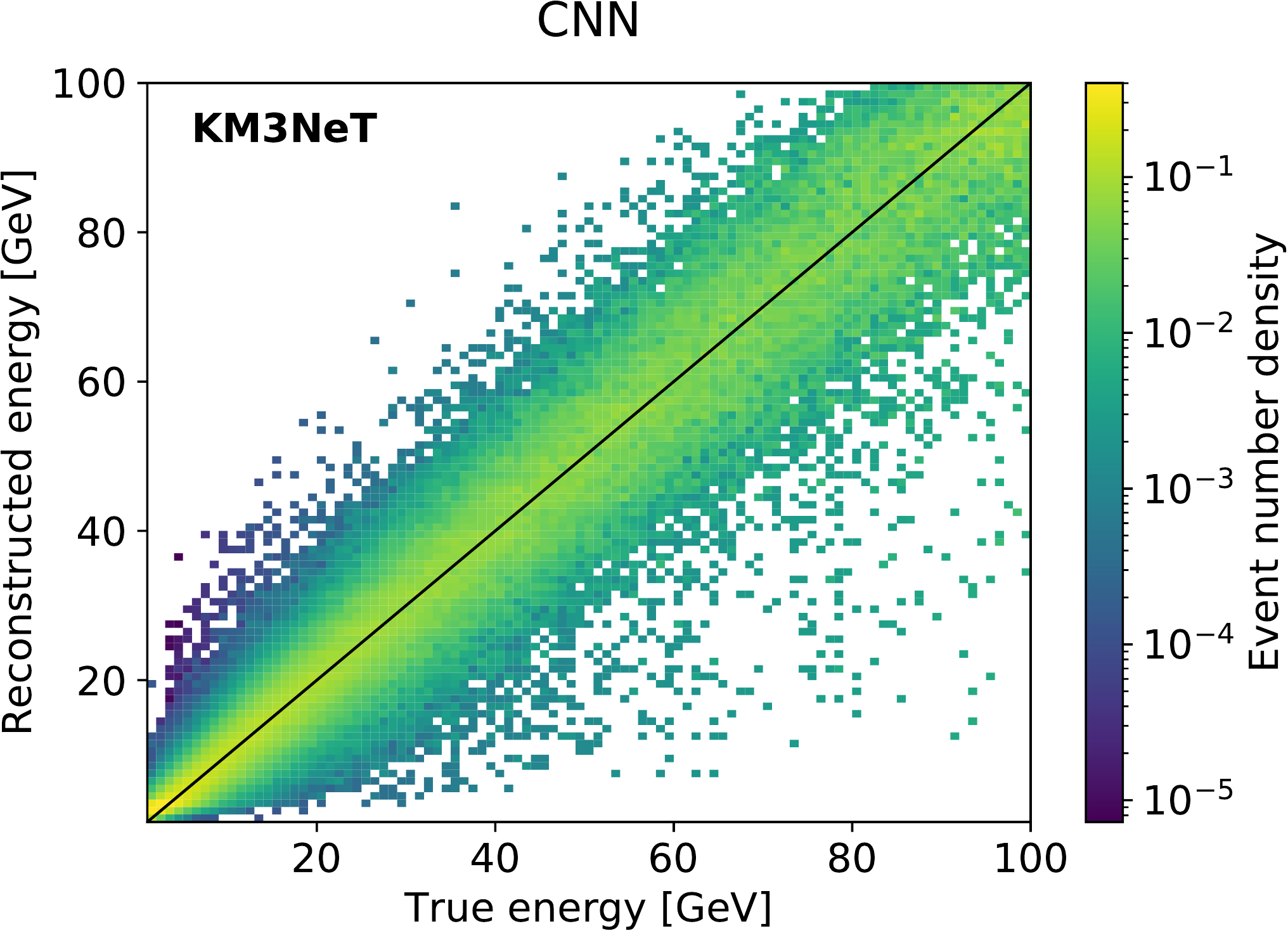}}
\end{minipage}
\hfill
\begin{minipage}[c]{0.48\textwidth}
\centering
    {\includegraphics[width =\textwidth, page=1]{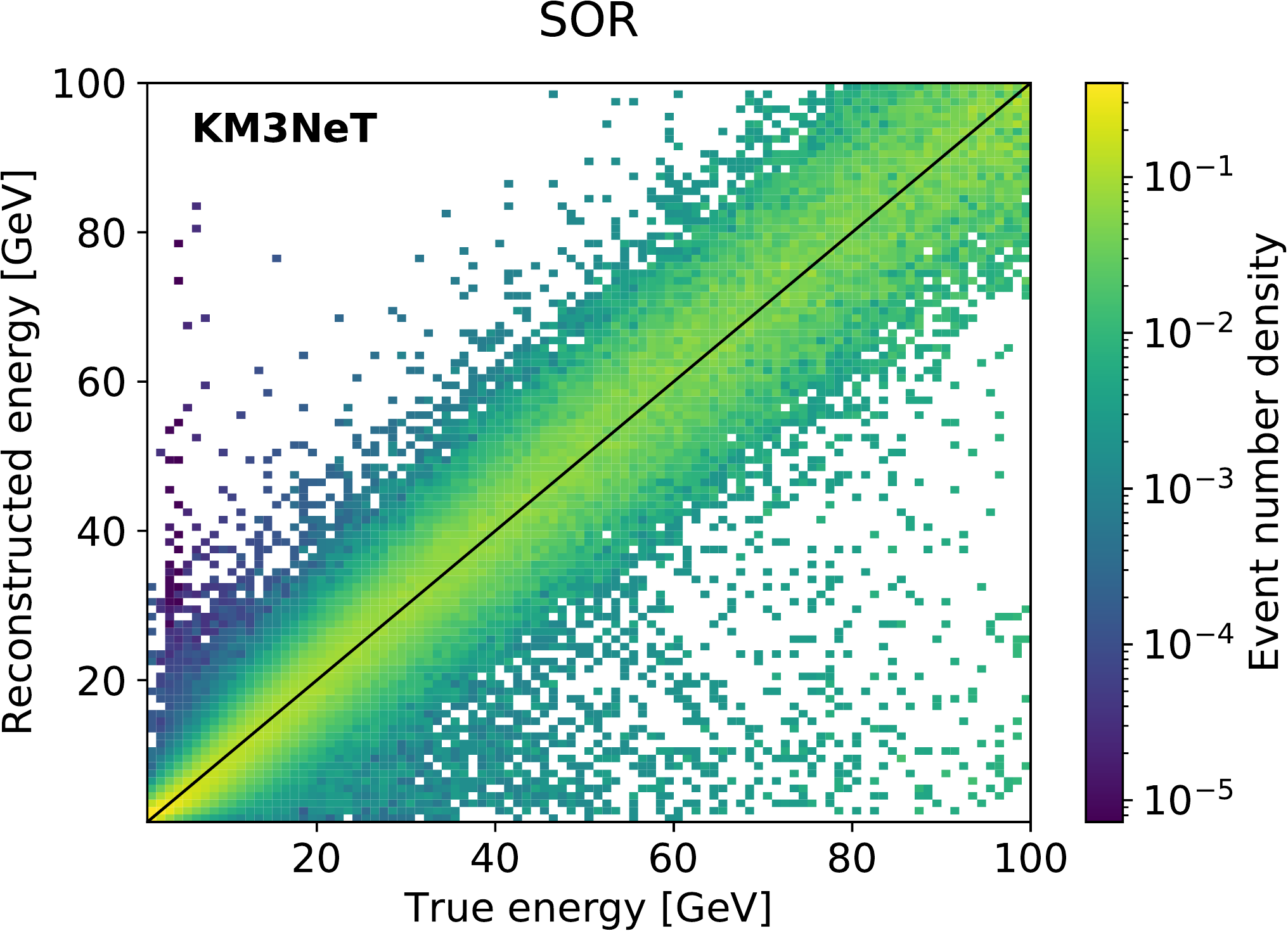}}
\end{minipage}
\caption{Reconstructed and corrected energy versus true MC neutrino
  energy for reconstructed $\nueCC$ events. The distribution has been
  normalised to unity in each true energy bin. Left: CNN event
  regressor. Right: Maximum-likelihood-based reconstruction algorithm
  SOR for shower-like events as described in
  Ref.~\autocite{Adrian-Martinez2016}.}
\label{fig:regr_2d_energy_nueCC}
\end{figure}

The reconstructed energy versus the true MC energy for reconstructed
$\nueCC$ events is shown in
\myfref{fig:regr_2d_energy_nueCC_not_corrected}.
As can be seen in the figure, the energy reconstruction tends to
underestimate the true MC energy of the neutrino, in particular for
energies above \SI{30}{\giga\electronvolt}. This effect has been found
to be even more evident for SOR, so that a correction function has
been applied to the reconstructed energy. This correction function has
been derived from MC simulations and depends on several reconstructed
quantities, such as the energy, the zenith angle, and the interaction
inelasticity (Sec.~D.1. in Ref.~\autocite{bib:hofestaedt2017}). So as to
allow for a comparison between the SOR algorithm and the CNN event
regressor, a similar but simpler correction function, only using the
reconstructed energy, has been derived and applied to the energy
reconstruction of the CNN event regressor for $\nueCC$ events. The
resulting distributions can be compared in
\myfref{fig:regr_2d_energy_nueCC}.
 
%
\noindent In order to allow for quantitative comparisons, the following two
performance metrics are used: the \textit{median relative error}
(MRE) and the \textit{relative standard deviation} (RSD). Both metrics
are calculated for each MC energy bin. The MRE is defined as:
\begin{equation}
\text{MRE} = \text{Median} \left( \left|
\frac{\text{E}_\text{reco} - \text{E}_\text{true}}{\text{E}_\text{true}} \right| \right)\,,
\end{equation}
where $\text{E}_\text{reco}$ ($\text{E}_\text{true}$) stands for the
reconstructed (true MC) energy.

\begin{figure}[h!]
\centering
\begin{minipage}[c]{0.48\textwidth}
\centering
    {\includegraphics[width=\textwidth, page=1]{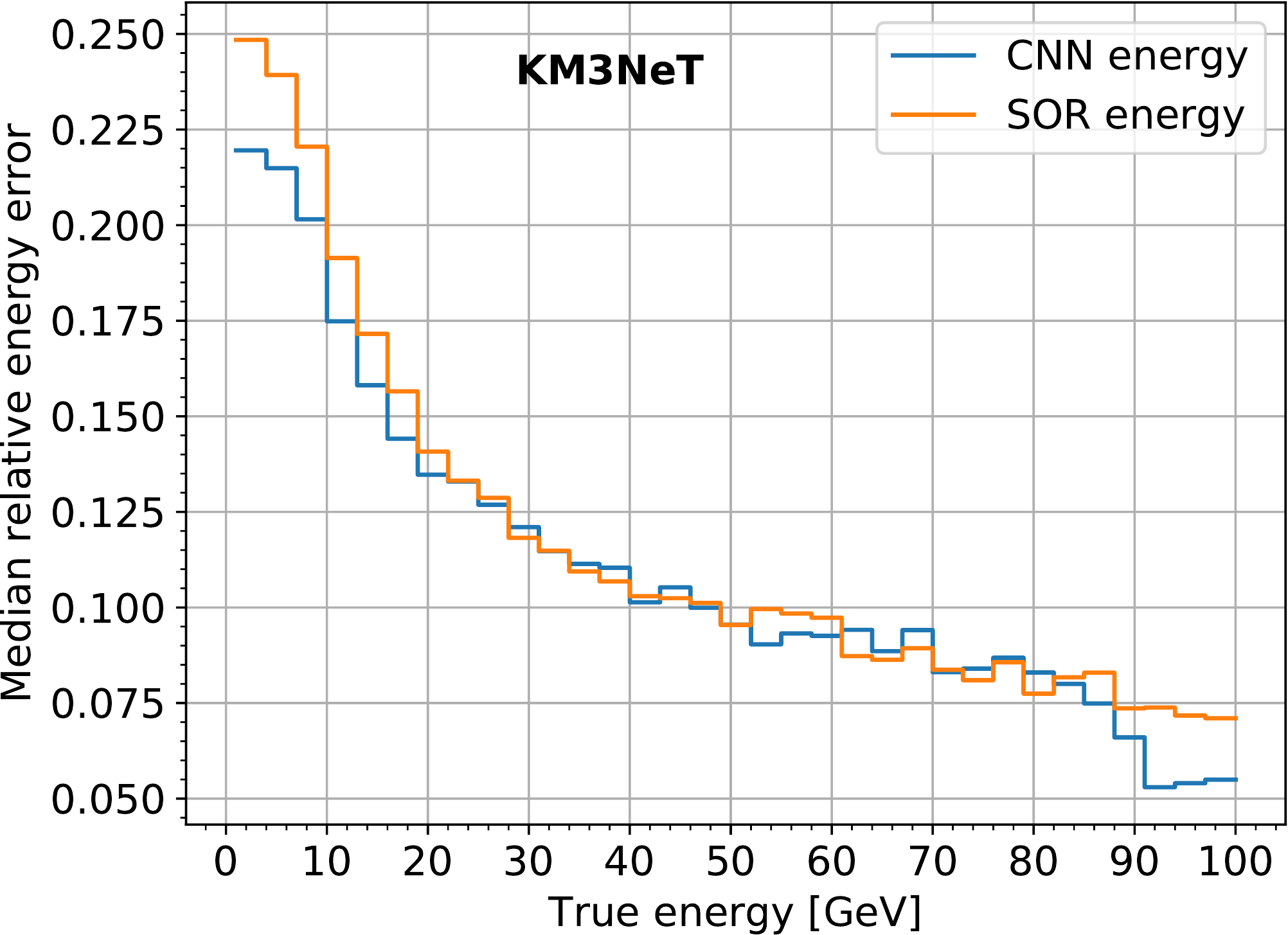}}
\end{minipage}
\hfill
\begin{minipage}[c]{0.48\textwidth}
\centering
    {\includegraphics[width =\textwidth, page=1]{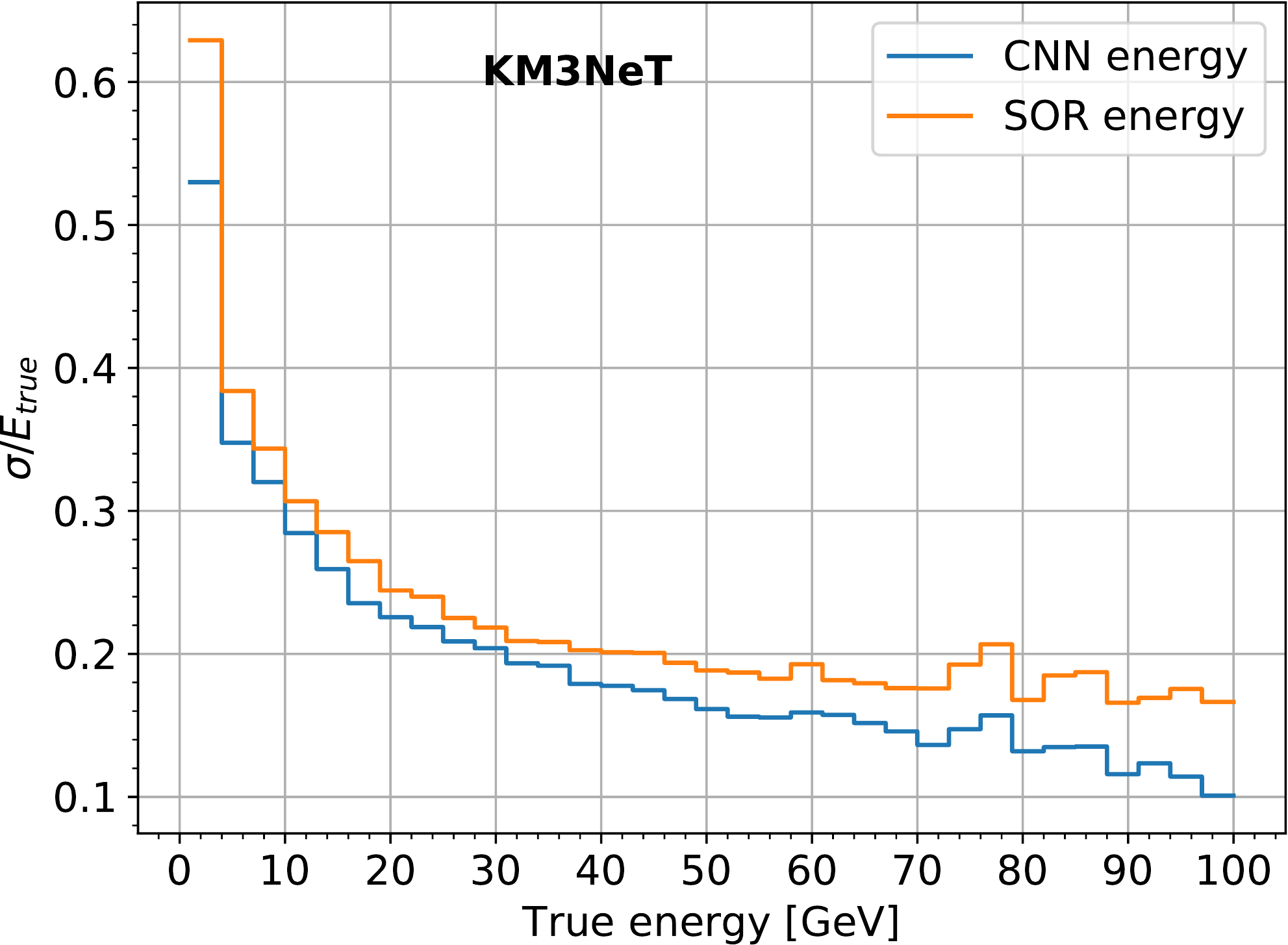}}
\end{minipage}
\caption{Left: Median relative error (MRE) on the reconstructed energy
  versus true MC energy for $\nueCC$ events for the CNN event
  regressor (blue) and SOR (orange). Right: Relative standard
  deviation (RSD) on the reconstructed energy versus true MC energy
  for the CNN event regressor (blue) and SOR (orange).}
\label{fig:regr_1d_energy_mre_rsd_nueCC}
\end{figure}

For the RSD, the standard deviation $\upsigma$ of the reconstructed
energy distribution for each MC true energy bin is calculated and then
divided by the energy of the bin:
\begin{equation}
\text{RSD} = \frac{\upsigma}{\text{E}_\text{true}}.
\end{equation}
For shower-like events, the energy reconstruction performance is
intrinsically limited by light yield fluctuations of the hadronic
particle cascade \autocite{Adrian-Martinez:2016zzs}.  Assuming that
both the CNN event regressor and the SOR algorithm come close to this
intrinsic resolution limit,
it is to be expected that the MRE and RSD
performances for both methods agree very well, as can be seen in
\myfref{fig:regr_1d_energy_mre_rsd_nueCC} (left) and
\myfref{fig:regr_1d_energy_mre_rsd_nueCC} (right), respectively.  For
the MRE, the CNN shows lower values than the SOR algorithm for
energies close to the boundaries of the investigated energy
range. This is due to the fact that the CNN learns the limits of the
simulated neutrino energy spectrum, i.e. \SI{1}{\giga\electronvolt}
and \SI{100}{\giga\electronvolt}. Therefore, the CNN, contrary to the
SOR algorithm, never reconstructs energies outside of this energy
range, which biases the MRE for the CNN towards lower values. For the
high-energy boundary, this effect can be easily avoided by extending
the simulated energy range. For the low-energy boundary, this is not
easily possible and biases the comparison with the SOR algorithm,
albeit only in an energy range that is close to the detection energy
threshold of KM3NeT/ORCA.  On the other hand, it is notable that the
RSD for the CNN event regressor improves by a few percent with respect
to the SOR algorithm also for energies far from the boundaries.
%
%
%
In the case of $\nueNC$ events, the energy reconstruction performance
is limited by the non-detectable energy that is carried away by the
outgoing neutrino.  For $\nueNC$ events, as for charged-current
events, the energy reconstruction performance of the CNN event
regressor and SOR is similar.
%
%
\subsection{Direction reconstruction performance}
The direction reconstruction performance of the CNN event regressor is
investigated in the same way as for the energy reconstruction. The X,
Y and Z components of the reconstructed direction vector are converted
to spherical azimuth and zenith coordinates. Since the DUs,
cf.~\mysref{subsect:detector_layout}, are approximately vertical
structures, the azimuth angle is defined by the projection of the
direction vector onto a plane roughly perpendicular to the DUs, while
the zenith roughly corresponds to the angle enclosed by the direction
vector pointing back to the particle's source and the DUs.

\begin{figure}[h!]
	\centering
        \includegraphics[width=0.675\textwidth, page=1]{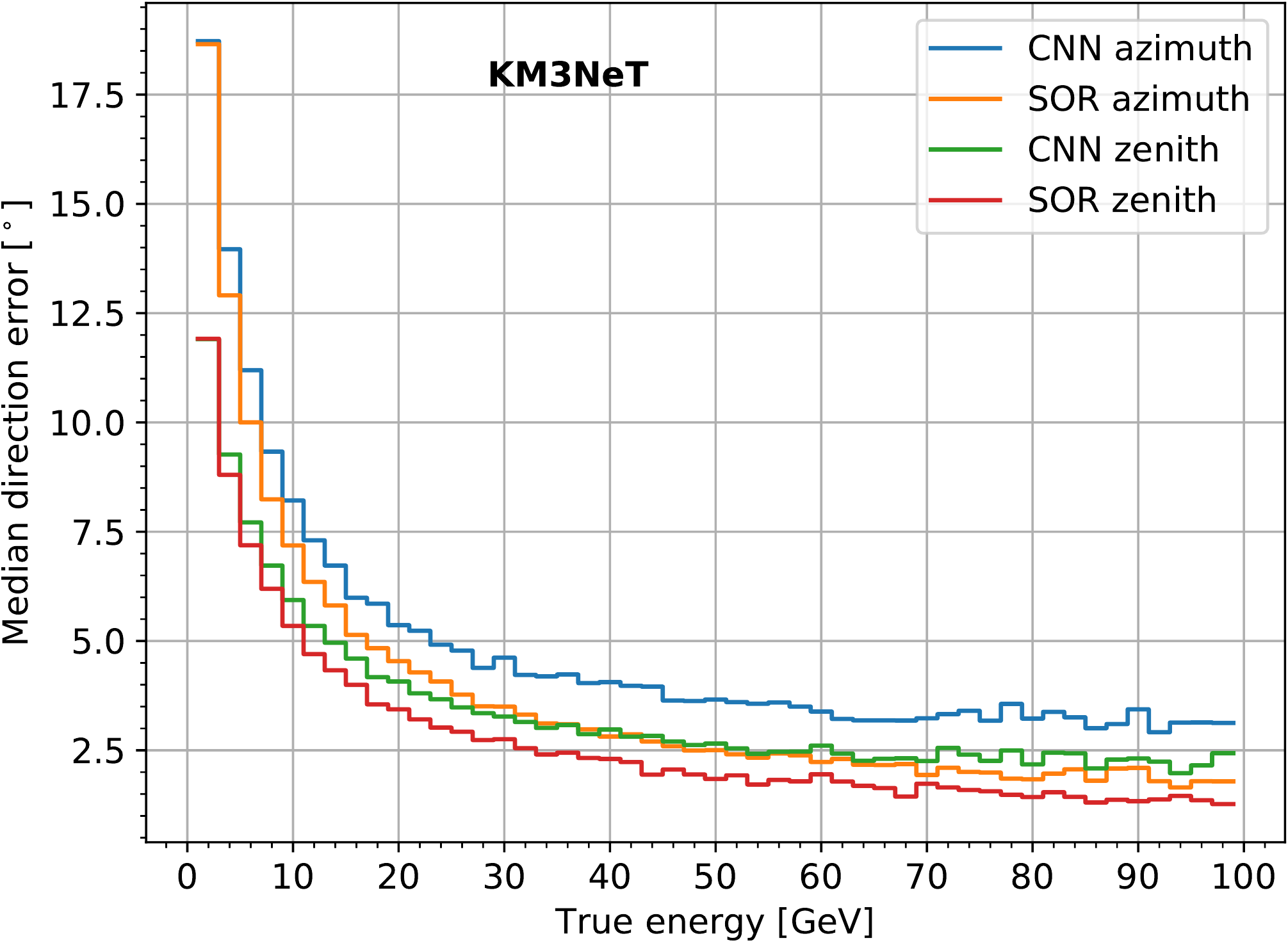}
  \caption{Median absolute error of the zenith and
    azimuth angle reconstruction for the CNN event regressor and the
    shower reconstruction algorithm SOR versus true MC
    energy for $\nueCC$ events.}
  \label{fig:regr_1d_zenazi_nuecc_median}
\end{figure}

%
%
The median absolute error (ME) is defined as the median of the
distribution of $\yreco - \ytrue$.  In
\myfref{fig:regr_1d_zenazi_nuecc_median}
the ME for the azimuth and
the zenith angle of the reconstructed direction is compared for the
CNN event regressor and SOR. The comparison is done for $\nueCC$
events and as a function of the true MC neutrino energy.
For energies below \SI{3}{\giga\electronvolt}, the ME is very similar
for both reconstructions, while the difference in performance
increases in favour of the SOR algorithm with increasing energy. At
\SI{20}{\giga\electronvolt}, the ME for the reconstruction of the
neutrino zenith (azimuth) angle is about 15\% (25\%) larger for the
CNN event regressor than for SOR. The reason for this worse
performance of the CNN might be connected to the rather coarse time
binning chosen for the image generation.
%
%
The corresponding comparison for $\nueNC$ events shows the same
general trend, but the performance gap is smaller since the
reconstruction is intrinsically limited by the interaction kinematics.
\subsection{Vertex reconstruction performance}
For the comparison of the vertex reconstruction performance the
distance between the reconstructed and the true MC vertex point is
investigated based on the longitudinal and perpendicular components of
the residual vector with respect to the direction of the incoming
neutrino.
%
The CNN event regressor is trained such that it reconstructs
the neutrino interaction vertex, while the SOR algorithm reconstructs
the brightest point of the Cherenkov light emission, which is shifted
by an energy-dependent displacement of the order of meters.

\begin{figure}[h!]
\centering
\begin{minipage}[c]{0.48\textwidth}
\centering
    {\includegraphics[width=\textwidth, page=1]{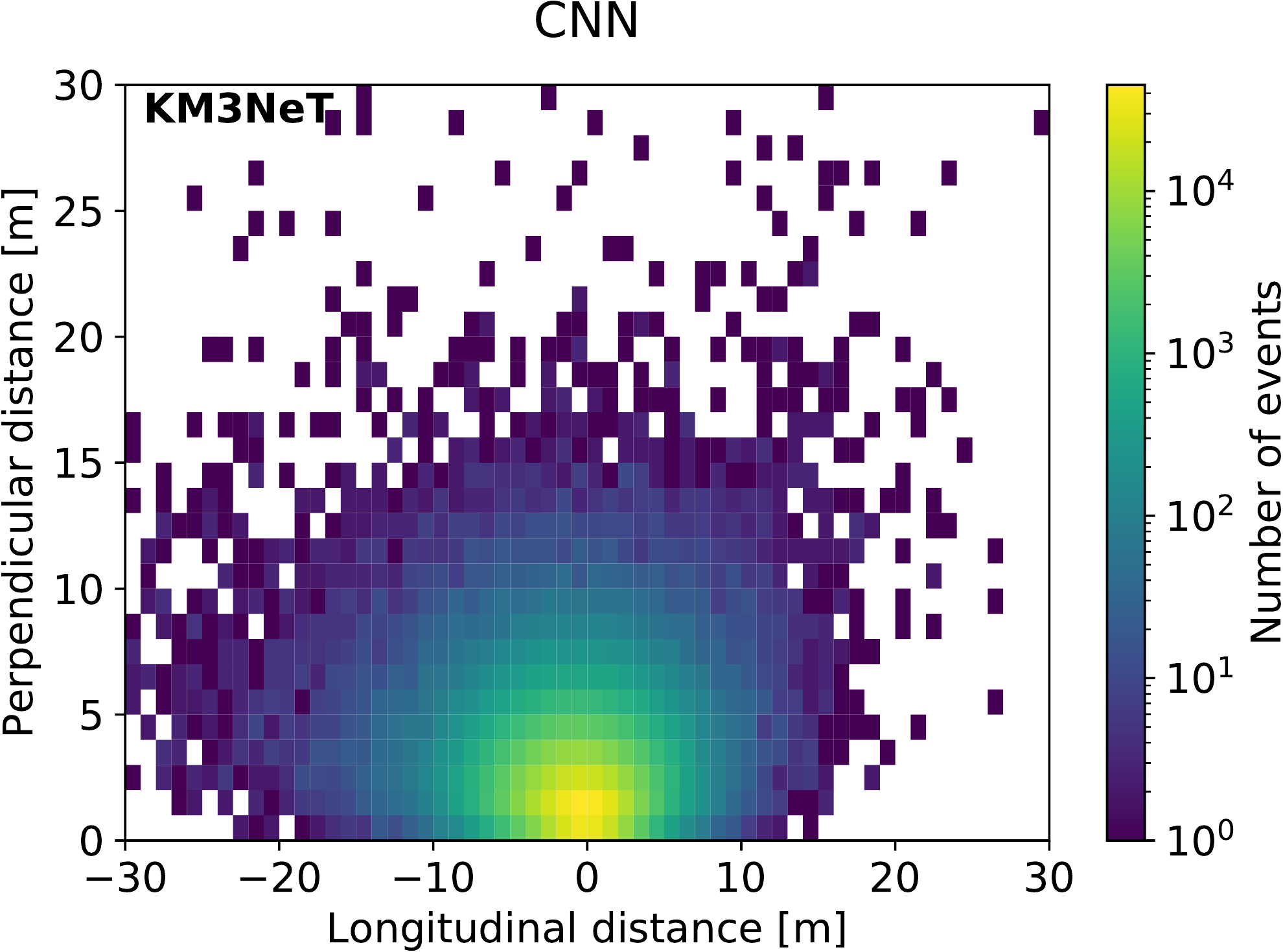}}
\end{minipage}
\hfill
\begin{minipage}[c]{0.48\textwidth}
\centering
    {\includegraphics[width =\textwidth,page=1]{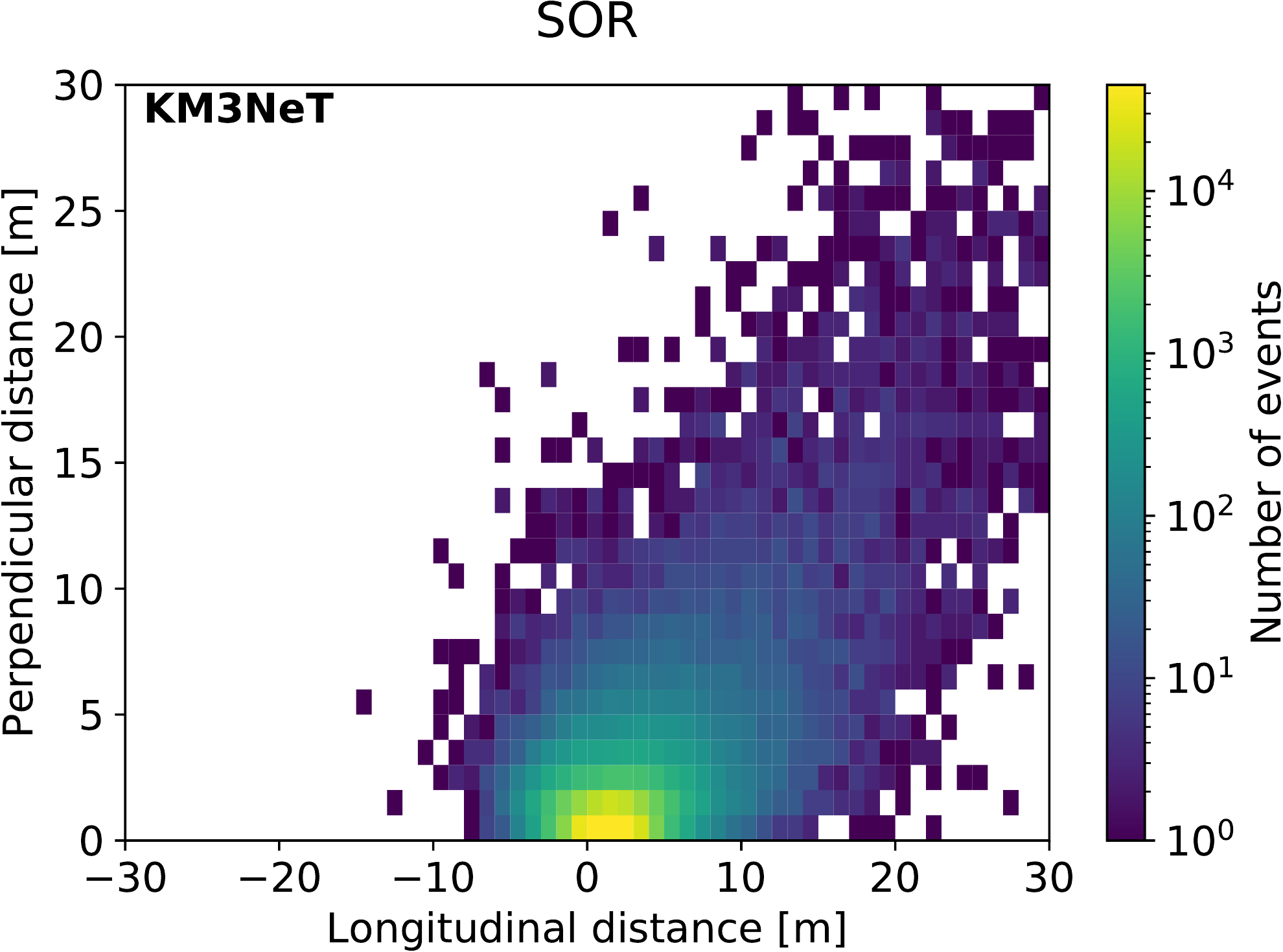}}
\end{minipage}
\caption{Reconstructed neutrino interaction point with respect to the
  true MC vertex for $\nueCC$ events. Shown is the perpendicular
  component versus the longitudinal component of the distance
  vector. Left: CNN event regressor; Right: SOR algorithm.}
\label{fig:regr_2d_vertex_nu_e_cc}
\end{figure}

The perpendicular versus longitudinal distance of the reconstructed
vertex with respect to the true $\nueCC$ interaction point is shown in
\myfref{fig:regr_2d_vertex_nu_e_cc} (left) for the CNN event
regressor, while the right plot shows the same distribution for the
SOR algorithm. The above mentioned displaced reconstruction of the
brightest emission point can be seen for the SOR algorithm in
\myfref{fig:regr_2d_vertex_nu_e_cc} (right). The CNN-based vertex
resolution is of the order of one meter with a small offset in
perpendicular direction in the investigated energy range up to
\SI{100}{\giga\electronvolt}. For the SOR algorithm, the resolution in
the longitudinal direction is similar, while it is better in the
perpendicular direction.  A correlation of longitudinal and
perpendicular errors can be seen for the SOR algorithm, which is
absent for the CNN reconstruction.
%
%
\subsection{Error estimation}

The CNN is used to reconstruct the regression uncertainty of all
reconstructed quantities, as explained in
\mysref{sec:regression_arch_loss}. For the following
performance investigation the event pre-selection, based on the
current KM3NeT/ORCA reconstruction pipeline, is not applied, i.e. all
triggered events of the simulated test dataset are used.
In order to evaluate the goodness of the uncertainty reconstruction
for any event quantity, e.g. energy or direction, the events are
binned according to their estimated uncertainty value,
$\upsigma_{\text{reco}}$. For the event distribution in each bin, the
standard deviation, $\upsigma_{\text{true}}$, of the residuals of the
reconstructed and true MC values $\left|\yreco - \ytrue\right|$ is calculated.

\begin{figure}[h!]
\centering
\begin{minipage}[c]{0.48\textwidth}
\centering
    {\includegraphics[width=\textwidth, page=1]{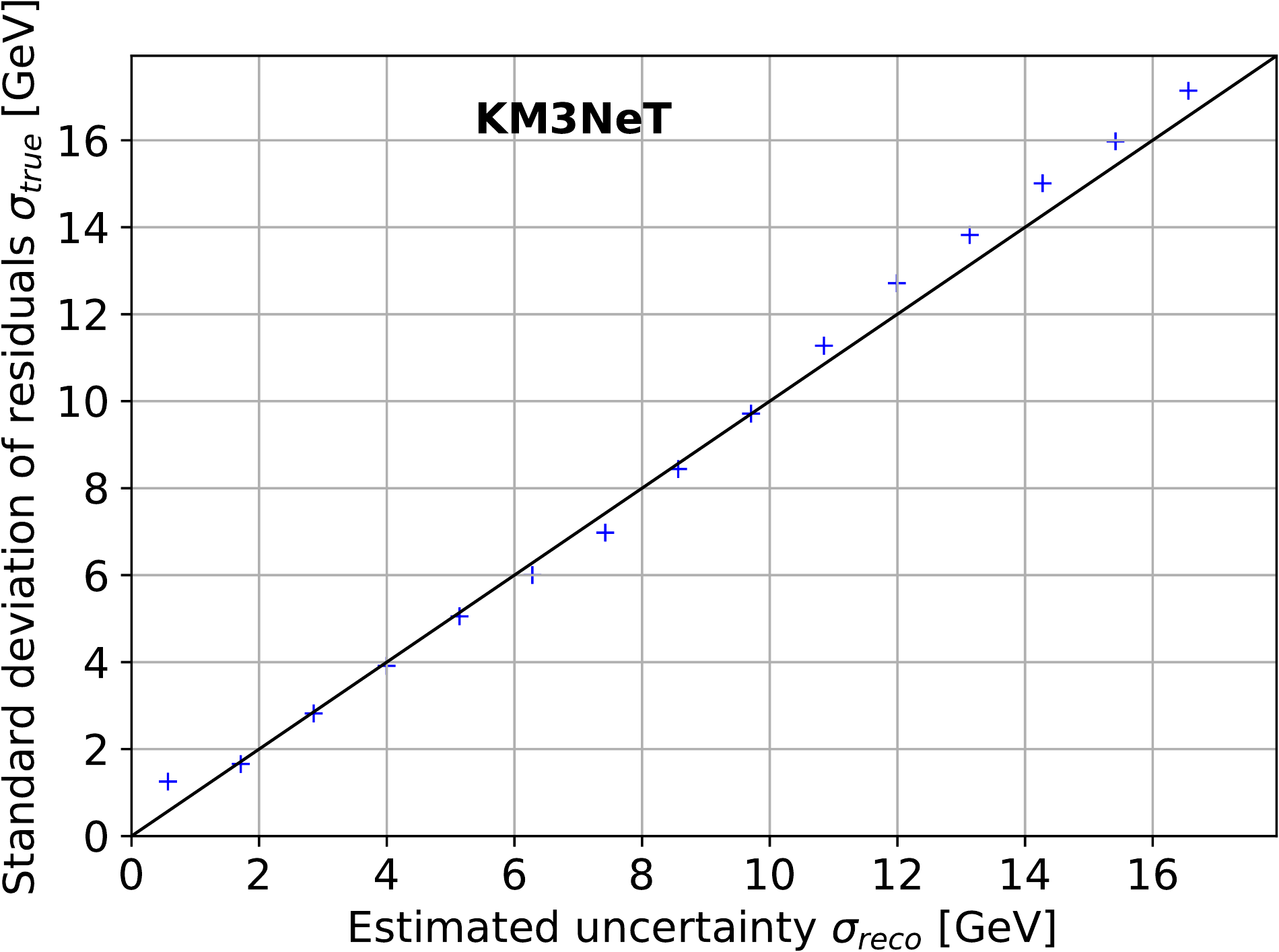}}
\end{minipage}
\hfill
\begin{minipage}[c]{0.48\textwidth}
\centering
    {\includegraphics[width =\textwidth,page=1]{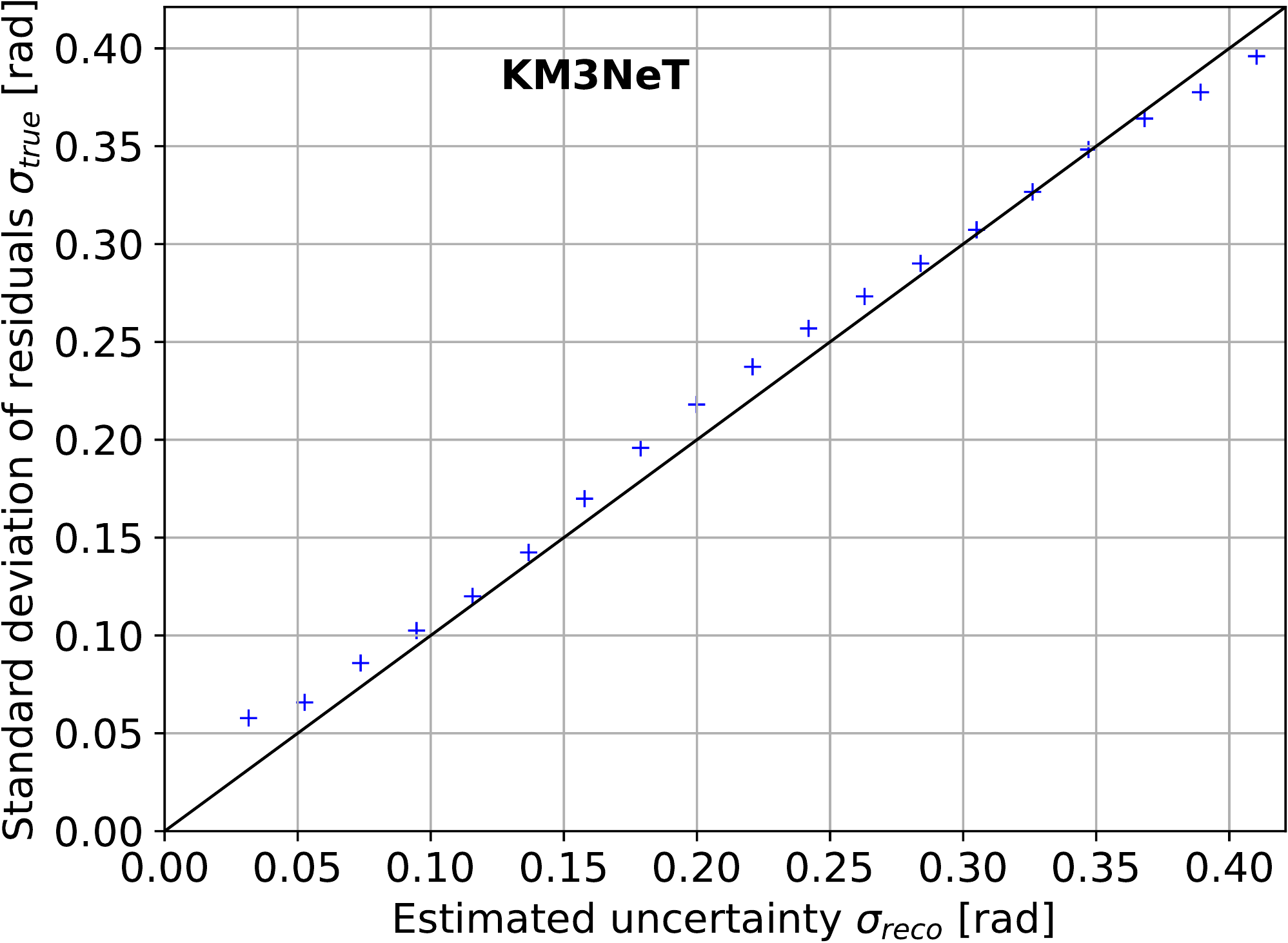}}
\end{minipage}
\caption{Standard deviation of the difference of true MC and
  reconstructed values versus the uncertainty estimated by the CNN
  event regressor. Left: for the energy reconstruction of $\nueCC$
  events. Right: for the Z component of the reconstructed direction
  vector for $\nueCC$ events.}
\label{fig:regr_2d_error_sigma_nu_e_cc}
\end{figure}

The distribution of $\upsigma_{\text{true}}$
versus $\upsigma_{\text{reco}}$ is shown in
\myfref{fig:regr_2d_error_sigma_nu_e_cc} (left) for the energy
reconstruction of $\nueCC$ events. Even though
$\upsigma_{\text{reco}}$ is estimated with a significant fraction of
events that were discarded previously by the criteria applied for the
energy reconstruction in \mysref{sec:regression_energy}, the CNN event
regressor estimation of the reconstruction uncertainty is clearly
correlated to the true uncertainty.  As can be seen, the energy
uncertainty is slightly underestimated for events that are difficult
to reconstruct and therefore show large estimated uncertainties.

The distribution of $\upsigma_{\text{true}}$ versus
$\upsigma_{\text{reco}}$ is shown in
\myfref{fig:regr_2d_error_sigma_nu_e_cc} (right) for the Z component
of the reconstructed direction vector in $\nueCC$ events.
In the range of $\text{0.15} < \upsigma_{\text{reco}} < \text{0.30}$ a slight
underestimation in the uncertainty reconstruction can be seen.
%
%
%

\begin{figure}[h!]
	\centering
        \includegraphics[width=0.675\textwidth,page=1]{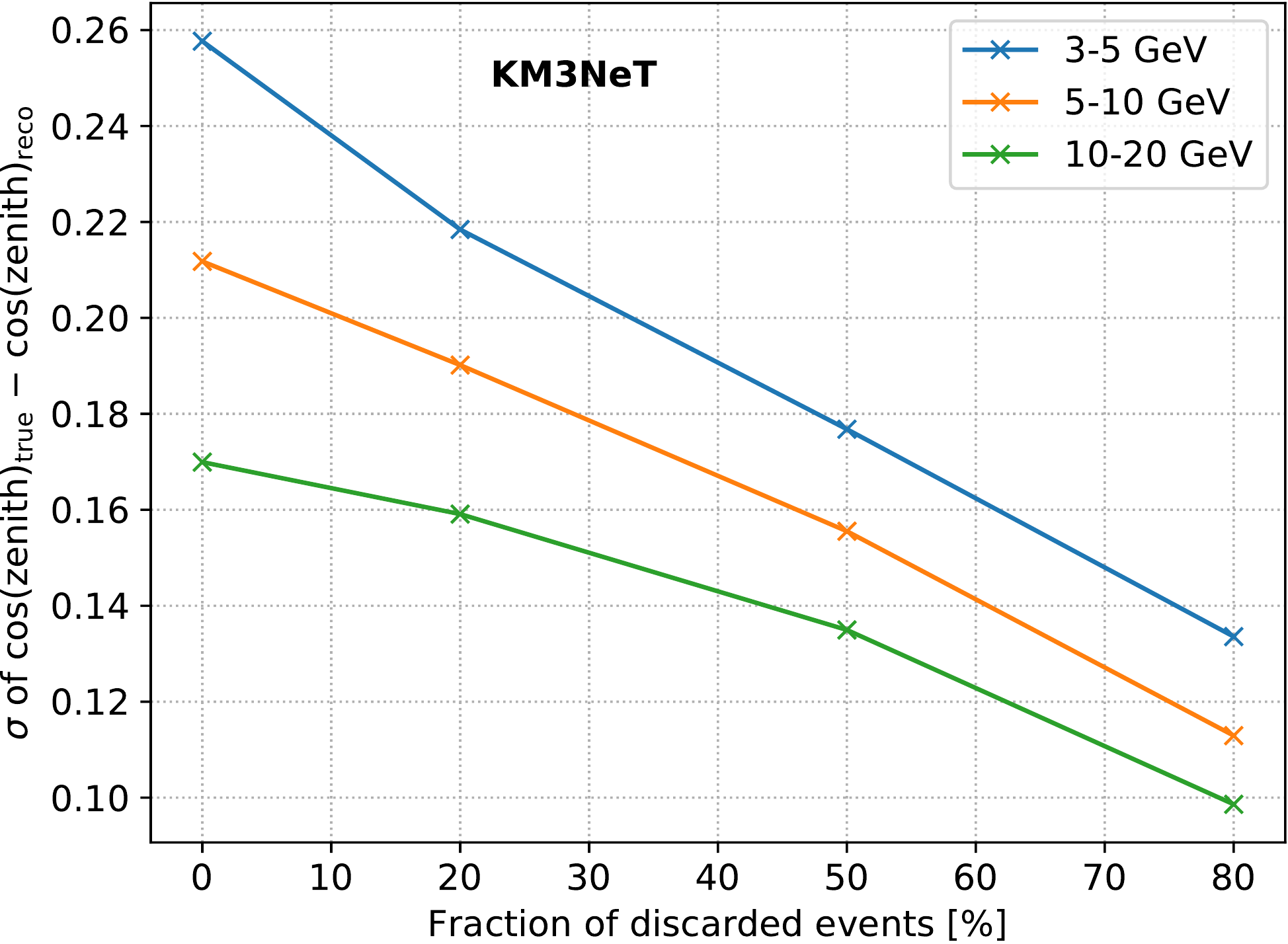}
	\caption{Standard deviation of the residual distribution of
            the reconstructed cosine of the zenith angle for $\nueCC$
            versus the fraction of events discarded by the CNN
            uncertainty estimator. The neutrino events have been
            generated in three different energy intervals as indicated
            by the three coloured lines.}
	\label{fig:regr_frac_discarded_events}
\end{figure}

\noindent Since there is a clear correlation between the predicted and the true
uncertainty, it is possible to discriminate badly reconstructed events
based on the estimated uncertainty value $\upsigma_{\text{reco}}$.
In order to demonstrate this, the zenith-angle reconstruction for
shower-like events is investigated.
The standard deviation
of the residual distribution of the reconstructed cosine of the zenith
angle for $\nueCC$ events versus the fraction of events discarded by
the CNN uncertainty estimator is depicted in \myfref{fig:regr_frac_discarded_events}.
For all three energy intervals
depicted, one can see that the zenith-angle resolution improves
significantly when the events are discarded,
that have the largest
reconstruction errors as predicted by the CNN.
%


\section{Conclusion}

In this work, the first application of deep convolutional neural
networks to the reconstruction and classification of simulated
neutrino events in the neutrino detector KM3NeT/ORCA has been
reported. Detailed MC datasets have been suitably pre-processed to
generate pixelated high-dimensional image input for the training of
CNNs. CNNs have been designed to separate neutrino
events from background, and track-like from shower-like neutrino
events. In addition, CNNs have been used to reconstruct the direction,
energy, and interaction point of the neutrinos, and to estimate the
uncertainty on each of the reconstructed quantities.  Comparisons to
the performance of classical machine-learning approaches to
classification and maximum-likelihood reconstructions show the
competitive and in some cases already superior performance of the
employed CNNs,
though there are still many aspects that can possibly
be optimised. In the next step, the developed CNNs will be tested for
robustness with respect to real  data acquisition conditions,
e.g.  for the effect of missing optical modules and imperfect
calibrations, before they will be applied to data taken with the
KM3NeT/ORCA detector.

\acknowledgments

The authors gratefully acknowledge the compute resources and support
provided by the Erlangen Regional Computing Center (RRZE).
Significant parts of this work are based on the soon to be published
doctoral thesis "Sensitivity studies on tau neutrino appearance with
KM3NeT/ORCA using Deep Learning Techniques" by M.~Moser.

The authors acknowledge the financial support of the funding agencies:
Agence Nationale de la Recherche (contract ANR-15-CE31-0020),
Centre National de la Recherche Scientifique (CNRS), 
Commission Europ\'eenne (FEDER fund and Marie Curie Program),
Institut Universitaire de France (IUF),
LabEx UnivEarthS (ANR-10-LABX-0023 and ANR-18-IDEX-0001),
Paris \^Ile-de-France Region,
France;
Shota Rustaveli National Science Foundation of Georgia (SRNSFG, FR-18-1268),
Georgia;
Deutsche Forschungsgemeinschaft (DFG),
Germany;
The General Secretariat of Research and Technology (GSRT),
Greece;
Istituto Nazionale di Fisica Nucleare (INFN),
Ministero dell'Universit\`a e della Ricerca (MUR),
PRIN 2017 program (Grant NAT-NET 2017W4HA7S)
Italy;
Ministry of Higher Education, Scientific Research and Professional Training,
Morocco;
Nederlandse organisatie voor Wetenschappelijk Onderzoek (NWO),
the Netherlands;
The National Science Centre, Poland (2015/18/E/ST2/00758);
National Authority for Scientific Research (ANCS),
Romania;
Ministerio de Ciencia, Innovaci\'{o}n, Investigaci\'{o}n y Universidades (MCIU): Programa Estatal de Generaci\'{o}n de Conocimiento (refs. PGC2018-096663-B-C41, -A-C42, -B-C43, -B-C44) (MCIU/FEDER), Severo Ochoa Centre of Excellence and MultiDark Consolider (MCIU), Junta de Andaluc\'{i}a (ref. SOMM17/6104/UGR), Generalitat Valenciana: Grisol\'{i}a (ref. GRISOLIA/2018/119) and GenT (ref. CIDEGENT/2018/034) programs, La Caixa Foundation (ref. LCF/BQ/IN17/11620019), EU: MSC program (ref. 713673),
Spain.



\end{document}